\documentclass[preprint,11pt]{aastex}
\usepackage{natbib}

\shorttitle{From infall to rotation around YSOs}
\shortauthors{Hogerheijde}

\begin{document}

\title{From infall to rotation around young stellar objects: A
transitional phase with a 2000~AU radius contracting disk?}

\author{Michiel R. Hogerheijde}

\affil{Radio Astronomy Laboratory, Astronomy Department, 601 Campbell
Hall, University of California, Berkeley, CA 94720-3411}

\email{michiel@astro.berkeley.edu}


\begin{abstract}
  
  Evidence for a transitional stage in the formation of a low-mass
  star is reported, intermediate between the fully embedded and the
  T~Tauri phases. Millimeter aperture synthesis observations in the
  HCO$^+$ $J$=1--0 and 3--2, HCN 1--0, $^{13}$CO 1--0, and C$^{18}$O
  1--0 transitions reveal distinctly different velocity fields around
  two embedded, low-mass young stellar objects.  The 0.6 M$_\odot$ of
  material around TMC~1 (IRAS 04381+2517) closely follows inside-out
  collapse in the presence of a small amount of rotation ($\sim 3$
  km~s$^{-1}$~pc$^{-1}$), while L1489~IRS (IRAS 04016+2610) is
  surrounded by a 2000 AU radius, flared disk containing
  0.02~$M_\odot$. This disk shows Keplerian rotation around a $\sim
  0.65$ $M_\odot$ star and infall at $1.3 (r/100\,{\rm AU})^{-0.5}$
  km~s$^{-1}$, or, equivalently, sub-Keplerian motions around a
  central object between 0.65 and $1.4$~$M_\odot$. Its density is
  characterized by a radial power law and an exponential vertical
  scale height. The different relative importance of infall and
  rotation around these two objects suggests that rotationally
  supported structures grow from collapsing envelopes over a few times
  $10^5$~yr to sizes of a few thousand AU, and then decrease over a
  few times $10^4$ yr to several hundred AU typical for T~Tauri
  disks. In this scenario, L1489~IRS represents a transitional phase
  between embedded YSOs and T~Tauri stars with disks. The expected
  duration of this phase of $\sim 5$\% of the embedded stage is
  consistent with the current lack of other known objects like
  L1489~IRS. Alternative explanations cannot explain L1489~IRS's large
  disk, such as formation from a cloud core with an unusually large
  velocity gradient or a binary companion that prevents mass accretion
  onto small scales.  It follows that the transfer and dissipation of
  angular momentum is key to understanding the formation of disks from
  infalling envelopes.

\end{abstract}



\section{Introduction\label{s:intro}}

A central question in the study of the formation of stars and
planetary systems is the dynamics of cloud core collapse, and the
relative importance of infall and rotation in particular. In the
general theoretical framework developed over the past 25 years
\citep[e.g.]{shu:selfsim,lizano:ambidiff,terebey:tsc,li:isopedic,basu:ambidiff1},
clouds slowly form dense, centrally concentrated cores which are
supported by a combination of turbulence, rotation, or weak magnetic
fields. These cores grow until they reach an unstable equilibrium,
after which they collapse on dynamical time scale, forming one or more
stars. As material from the collapsing core accretes onto the central
object(s), conservation of angular momentum leads to increased
flattening and the growth of a circumstellar disk. These disks,
detected around the majority of young stars
\citep{beckwith:mmdisks,beckwith:nature}, are the progenitors of
planetary systems. Little is known quantitatively about this
transition from an infalling core to a rotationally supported
disk. Early work by \citet{terebey:tsc} treated rotation as a
perturbation to collapse, and was therefore limited to regions where
rotation does not dominate the velocity field. Several groups have
addressed the formation of flattened, disk-like structures without
rotation but resulting from magnetic fields or initial conditions
\citep[e.g.]{galli:pseudo1,li:isopedic,hartmann:sheets}. Especially
relevant to the observations described here are the numerical
simulations by \citet{yorke:disks3} and \citet{nakamura:disklike}.

This paper investigates the velocity structure around two embedded
young stellar objects (YSOs: L1489~IRS and TMC~1) in detail, in order
to shed more light on the mechanisms and time scales by which disks
form from collapsing envelopes. Both YSOs are classified as Class I
objects, with bolometric temperatures \citep{chen:blt} of 238~K
(L1489~IRS) and 139~K (TMC~1), and bolometric luminosities of 3.7 and
0.7 $L_\odot$, respectively. In the near-infrared, they are visible in
light scattered through their outflow cavities
\citep{padgett:nicmos}, suggesting in both cases that we are viewing
the system at inclinations between $60^\circ$ and $90^\circ$.
Previous interferometric observations in HCO$^+$, $^{13}$CO, and
C$^{18}$O show flattened structures with indications of Keplerian
rotation \citep{mrh:taurus2,brown:rotenv}. Taken together, these
observations have led to a consensus picture, where the two YSOs are
embedded in collapsing envelopes which become increasingly flattened
and dominated by rotation on smaller ($\lesssim 1000$ AU) scales.

Recently, \citet[hereafter Paper~I]{mrh:scuba} analyzed the
submillimeter-continuum emission as observed with SCUBA on the JCMT of
four YSOs, including L1489~IRS and TMC~1. This work concluded that the
density distribution around all four YSOs follows radial power-laws
with indices between $-1$ and $-2$, consistent with simple inside-out
collapse \citep{shu:selfsim}. Their velocity structure as traced by
HCO$^+$ 3--2 and 4--3 line profiles confirmed that three of the four
envelopes were indeed undergoing inside-out collapse, including
TMC~1. The line profiles toward L1489~IRS did \emph{not} match the
velocities predicted by the inside-out collapse model for the best-fit
parameters obtained from its continuum emission. Instead, this model
predicted line widths exceeding the observations by factors of a
few. Paper~I concluded that L1489~IRS was instead surrounded by a
2000~AU radius, rotating, disk-like structure. The SCUBA data also
revealed a starless core $\sim 1'$ northeast of L1489~IRS, named
L1489~NE-SMM, indicating that truly isolated star formation is rare
even in Taurus.

The hypothesis that L1489~IRS is surrounded by a disk-like structure
is further investigated here, in particular in the framework of the
growth of a rotationally-supported disk within an infalling
envelope. Previous aperture-synthesis observations of these sources
already showed hints of the underlying dynamical structure, which was
interpreted in both cases as indicative of rotation on $\lesssim 1000$
AU scales.  This paper presents new observations in HCO$^+$ 1--0 and
3--2, $^{13}$CO 1--0, C$^{18}$O 1--0, and HCN 1--0, obtained with the
Berkeley-Illinois-Maryland Association array and with the Owens Valley
millimeter array. The new data greatly improve the sensitivity and the
coverage of different spatial scales, allowing a detailed,
quantitative analysis of the velocity and density structure of the
envelopes of both YSOs. We find that the velocity fields around both
objects are remarkably different, suggesting infall in the presence of
some rotation around TMC~1 and Keplerian rotation with some inward
motions in a 2000~AU radius disk around L1489~IRS.

The outline of the paper is as follows. Section \ref{s:obs} describes
the observations, and briefly discusses the methods used to combine
data obtained from the two millimeter arrays. Section \ref{s:results}
presents the resulting integrated-intensity images and
position-velocity diagrams, while section \ref{s:models} explores
models of the position-velocity diagrams and discusses the findings in
an evolutionary framework where disks grow from collapsing envelopes
to size of a few thousand AU and subsequently contract to several
hundred AU. Section \ref{s:conclusions} concludes the paper by
summarizing the main conclusions.

\section{Observations and data reduction\label{s:obs}}

Table \ref{t:obs} lists the dates of the observations presented in
this work.  The initial observations of L1489~IRS and TMC~1 with the
Owens Valley Millimeter Array\footnote{The Owens Valley Millimeter
Array is operated by the California Institute of Technology under
funding from the U.S.\ National Science Foundation (\#AST96--13717).}
(OVRO) took place in 1993--1997, and are described in detail in
\citet{mrh:taurus1,mrh:taurus2}. Observations to significantly
increase the sensitivity and expand the $uv$-coverage followed in
1998--2000, and are further discussed here.

The HCO$^+$ $J$=1--0 and HCN $J$=1--0 lines were observed using the
Berkeley-Illinois-Maryland Association \citep[BIMA]{welch:bima}
interferometer\footnote{The BIMA Array is operated by the
Berkeley-Illinois-Maryland Association under funding from the National
Science Foundation.} in 1998 and 1999. Two configurations of the ten
6-meter antennas were used: the B-array, with projected baselines
between 3 and 70~k$\lambda$, and the C-array (2--24~k$\lambda$). The
digital correlator recorded the two lines in 12.5~MHz wide bands with
256 channels each, giving in a velocity resolution of
0.16~km~s$^{-1}$. The two remaining windows of the correlator
registered the 3.4~mm continuum in a bandwidth of 400~MHz (upper +
lower side band). Phase and amplitude variations were calibrated by
observing the nearby quasars 3C84, 0238+166, and 0530+135
approximately every 30 minutes. The adopted fluxes of these quasars of
4.0, 3.0, and 3.0~Jy, respectively, were measured during the time of
the observations using planets as primary calibrators. The calibration
was performed using the MIRIAD (Multichannel Image Reconstruction,
Image Analysis and Display; \citealt{sault:miriad}) task GFIDDLE. In
addition to these single-pointed observations, three-point mosaics
covering the core $1'$ east of L1489~IRS and the outflow of TMC~1 in
HCO$^+$ and HCN 1--0 were obtained in a similar way with the BIMA
array in the C-configuration in 2000 July.

The $^{13}$CO $J$=1--1 and C$^{18}$O $J$=1--0 lines were observed with
the OVRO array in 1999 and 2000. The six 10-meter antennas were placed
in the low- and high-resolution configurations, with respective $uv$
coverage of 2--42 and 8--90~k$\lambda$. Each of the two lines were
recorded in two adjacent 2-MHz 32-channel wide bands with four
overlapping channels. This resulted in a total velocity coverage of
10~km~s$^{-1}$ with a velocity resolution of 0.16~km~s$^{-1}$. The
2.7~mm continuum was registered separately in a bandwidth of 1~GHz.
Phase and amplitude variations were calibrated using the MMA software
package specific to OVRO \citep{scoville:database}. The quasars 3C84
and 0333+321 served as gain calibrators with integrations every 20--30
minutes. Absolute flux calibration was obtained either from
observations of Uranus, or from observation of bright quasars at the
start or end of the track with fluxes measured during the period of
the observations: 3C84, $F=4.0$ Jy; 3C454.3, 7.2 Jy; and 3C273, 9.9
Jy. To further increase the sensitivity to extended emission, the same
$^{13}$CO and C$^{18}$O 1--0 lines were also observed with the BIMA
array in the C-configuration in 2000 July, with an observational setup
similar to that for the other BIMA observations.

On 2000 December 5 L1489~IRS was observed at BIMA in the HCO$^+$
$J$=3--2 transition. The antennas were placed in C-configuration with
projected baselines between 6 and 64~k$\lambda$. Of the nine antennas
with 1~mm receivers, only eight could be successfully tuned to this
high frequency; their nominal tuning range extends only to
270~GHz. The data from two more antennas were deleted because they had
system temperatures twice the typical value of $\sim 1200$~K. The
HCO$^+$ 3--2 line was recorded in a single correlator band with 256
98~kHz-wide (0.11~km~s$^{-1}$) channels and a total bandwidth of
25~MHz (28~km~s$^{-1}$). The continuum was recorded in a bandwidth of
275~MHz. Even under the favorable weather conditions during the
observations (rms of path length variations over a 100~m baseline of
181~$\mu$m; 8.8~mm of precipitable water), calibrating the phase
variations requires much care. Observations of the phase calibrator,
the quasar 3C84, track the total phase drifts which span several
hundred degrees over the course of the track with point-to-point
fluctuations of $\sim 70^\circ$ rms. However, it was found that the
gain solution found from 3C84 cannot be applied to L1489~IRS because
of additional atmospheric fluctuations over their projected distance
of $17^\circ$: Compared to the HCO$^+$ 1--0 image (see section
\ref{s:results}) it produces multiple copies of the object offset from
the field center by $5''$, and places considerable flux in the
sidelobes of the dirty beam. Instead we use the HCO$^+$ 1--0 data as a
model to self-calibrate the HCO$^+$ 3--2 observations. Using only the
$uv$ range of 6--64~k$\lambda$ in common with the 3--2 data, the image
of the HCO$^+$ 1--0 emission averaged over the velocity range of 6--9
km~s$^{-1}$ where most of the emission is found is almost
point-like. Because the 3--2 line traces even denser material than
1--0, presumably found closer to the object, using this HCO$^+$ 1--0
image as a model to self-calibrate the HCO$^+$ 3--2 data, averaged
over the same velocity range, is a conservative assumption. A
time-averaging interval of 20~minutes was used, giving a
signal-to-noise exceeding 6 is found on those baselines that have
significant emission. This is sufficient to base a self-calibration
solution on.  In strict terms this procedure means that the HCO$^+$
3--2 image is now no longer fully independent of the HCO$^+$ 1--0
data. However, the velocity structure is still independent
because we have used the emission averaged over all channels between 6
and 9 km~s$^{-1}$ for the self-calibration. The calibration of the
flux scaling was carried out by adopting a flux of 1.7~Jy for 3C84,
obtained by extrapolating recent observations at 90 and 230~GHz to
267~GHz. The uncertainty in the flux scaling is estimated at $\sim
50$\%.

Subsequent processing of the data, including the joining of the two
bands covering the $^{13}$CO and C$^{18}$O 1--0 lines in the 1999/2000
OVRO data, was carried out with the MIRIAD software package. The new
(1998--2000) and previously obtained (1993--1997) 3~mm line data were
combined by interpolating the visibilities on a common velocity grid
with a resolution of 0.17~km~s$^{-1}$. A more detailed description of
the steps involved in combining data from the BIMA and OVRO
instruments (the virtual CARMA\footnote{Combined Array for Research in
Millimeter Astronomy, a joint initiative of the University of
California at Berkeley, the University of Illinois at
Urbana-Champaign, the University of Maryland, and the California
Institute of Technology.} array) is given in Hogerheijde et al.\
(in prep.). Since
the primary beam size of the instruments differs ($86''$ for OVRO
vs. $143''$ for BIMA at 3.4~mm; $68''$ vs. $114''$ at 2.7~mm), only
emission within the smaller of the two fields is properly sampled in
the combined data. Since our main interest is in the emission in the
inner few thousand AU of the envelopes, this is of no immediate
concern here. Only the three-point mosaics address the larger scale
emission, but they were obtained with the BIMA array alone and do not
suffer from this effect. The model calculations presented in \S
\ref{s:models} take these different primary beam rigorously into
account.

Images were produced using MIRIAD's CLEAN algorithm and `robust'
weighting (robustness parameter 0--2) of the visibilities to optimize
the signal-to-noise and the spatial resolution. In some cases
(L1489~IRS: C$^{18}$O 1--0; TMC~1: $^{13}$CO and C$^{18}$O 1--0), the
images were also convolved with a $3''$ Gaussian to further improve
the signal-to-noise. Resulting noise levels are 0.1--0.3
Jy~beam$^{-1}$ (1--3 K) in 0.17~km~s$^{-1}$ channels for the 3~mm line
emission, 0.3 mJy~beam$^{-1}$ in a 0.11~km~s$^{-1}$ channel at 1~mm;
and 0.9--1.3 mJy~beam$^{-1}$ at 3~mm and 70 mJy~beam$^{-1}$ at 1~mm
for the continuum images; typical beams have a FWHM of $3''$--$6''$
(Tables \ref{t:cont} and \ref{t:line}). Integrated-intensity and
velocity-centroid images were obtained from the cleaned spectral-line
cubes using a $3\sigma$ clip level. The HCO$^+$ 1--0 and HCN 1--0
three-point mosaics were deconvolved with the MIRIAD task MOSMEM, a
maximum entropy method better suited to handle extended structures and
mosaics.

\section{Results\label{s:results}}

\subsection{Continuum emission\label{s:cont}}

Continuum emission at 3.4 and 2.7~mm is detected toward both YSOs, at
levels of 5--10~mJy depending on source and frequency
(Fig. \ref{f:cont} and Table \ref{t:cont}). No emission is detected at
267~GHz (1.1~mm) to L1489~IRS, with a 2$\sigma$ upper limit of
70~mJy. The positions listed in the Table agree within $2''$ with
those inferred from the single-dish submillimeter continuum
measurements of Paper~I, a good match given the spatial resolution in
both data sets and the relatively low signal-to-noise of the
interferometric 3~mm continuum data. This signal-to-noise ratio does
not allow conclusions about the spatial distribution of the emission,
e.g., using the amplitudes as function of $uv$ distance
(Fig. \ref{f:cont}).

Still, limits can be placed on the contribution from unresolved
sources to the continuum flux by using the models of Paper~I. The
best-fit envelope models of Paper~I predict fluxes at 2.7~mm of
3.0~mJy (L1489~IRS) and 1.1~mJy (TMC~1).  These fluxes follow from
Fourier-transformation of the model images, sampling at the $(u,v)$
positions of the data, transformation back to the image plane, and
deconvolution. The resolving-out of extended emission is therefore
accurately accounted for. The resulting fluxes are of the same order
as those observed (5.8 and 7.7 mJy, respectively). A similar value of
3~mJy is found for L1489~IRS using the model developed in \S
\ref{s:l1489}, while the upper limit of 70~mJy at 1.1~mm is consistent
with the predicted flux at 1.1~mm of 50~mJy. When taking into account
that the models may underestimate the dust temperature close to the
source (see Paper~I), no more than a few mJy is left for any
additional, unresolved component. Adopting a dust temperature of 30~K
and a `standard' dust emissivity value at 2.7~mm of 1~cm$^{2}$~g$_{\rm
dust}^{-1}$ \citep[e.g.]{ossenkopf:kappa,pollack:kappa}, these flux
limits translate to gas+dust masses of $\sim 10^{-3}$ $M_\odot$,
providing a strict upper limit to any compact dust disk around
L1489~IRS or TMC~1 in addition to the more extended distribution of
material modeled in \S \ref{s:models}. Centimeter-wave observations
\citep{rodriguez:radiocores,lucas:radio} suggest that no more than
0.5~mJy can be attributed to free-free emission at 3~mm.

\subsection{Line emission\label{s:lines}}

The incomplete sampling of extended emission by interferometric
observations of embedded YSOs often plays an important role in the
character of deconvolved images, and our HCO$^+$, HCN, $^{13}$CO, and
C$^{18}$O 1--0 observations of L1489~IRS and TMC~1 are no
exception. The HCO$^+$ 3--2 observations of L1489~IRS are less
affected by incompletely sampled large scale emission, because the
higher critical density of this transition limits the emission to
smaller regions and because fewer short $uv$ spacings are present in
the data set, filtering out extended emission more efficiently.  Fig.\
\ref{f:l1489_robust} shows the integrated emission in HCO$^+$ 1--0
toward L1489~IRS, comparing the uniformly-weighted image to the
naturally weighted image. These two schemes represent the extremes of
visibility weighing when inverting to the image plane: Natural
weighting assigns equal weight to all visibilities, resulting in a
lower resolution, but also lower noise and increased sensitivity to
extended emission. Uniform weighting assigns equal weight to each
interval in the $(u,v)$-plane, effectively down-weighing the more
densely sampled short baselines. This gives a higher resolution, but
increases the noise, and only small-scale structures are
recovered. These weighting schemes produce dramatically different
images of L1489~IRS in HCO$^+$ 1--0.  With uniform weighting only a
thin elongated structure is recovered at the position of the
source. For natural weighting, this structure still dominates the
image, but has become more bloated, and extended emission fills the
northeast section of the image. The size of the emitting region is
bound by the half-power primary beam width of the OVRO antennas
(having the smaller primary beam of the combined BIMA and OVRO data)
at $\sim 60''$ from the field center. Negative `emission' is present
in the southern half of the image, indicating that significant amounts
of extended flux are missing from the data.  Comparison with
single-dish observations of HCO$^+$ 1--0 in the $28''$ beam of the
IRAM~30m antenna \citep{mrh:taurus1} indicates that from the central
half arcminute all flux is recovered. Since it is this central region
to which most of our attention is focussed, the missing larger scale
flux is of no immediate concern.

Figure \ref{f:vcen} show the integrated-intensity images (contours) of
all observed lines toward L1489~IRS and TMC~1, superposed on the
velocity-centroid images (color scale; see below). The velocity planes
of the data cubes were individually `cleaned', after which the
integrated-intensity (zeroth moment) and velocity-centroid (first
moment) images were constructed using a clip-level of 3$\sigma$;
negative emission was not included in the velocity-centroid images for
clarity. All images are optimized for angular resolution and
sensitivity using `robust' weighting \citep{briggs:robust}, with a
`robustness parameter' dependent on the transition and source. This
parameter can vary between $+2$ and $-2$, where $+2$ approaches
natural weighting and $-2$ uniform weighting. For L1489~IRS, this
parameter was 0 for HCO$^+$ 1--0 and 2 for HCO$^+$ 3--2, 2 for HCN, 0
for $^{13}$CO, and 2 for C$^{18}$O; in addition, the C$^{18}$O image
was smoothed with a $3''$ Gaussian to increase signal-to-noise. For
TMC~1, the robustness parameter was 2 for all transitions; the
$^{13}$CO and C$^{18}$O images were also smoothed with a $3''$
Gaussian. This resulted in synthesized beam sizes with FWHM between
$3''$ and $5''$, as shown in the lower left of the panels of Fig.\
\ref{f:vcen}. For the velocity-centroid images, the adopted systemic
velocity of 7.3~km~s$^{-1}$ for L1489~IRS and 5.2~km~s$^{-1}$ for
TMC~1 are based on single-dish observations of optically thin tracers
such as C$^{18}$O and H$^{13}$CO$^+$ \citep{mrh:taurus2}.

The emission around L1489~IRS (Fig. \ref{f:vcen}) shows the
elongated structure around the YSO with a red/blue velocity structure
along the major axis indicative of rotation, earlier noted by
\citet{mrh:taurus2}. In the HCO$^+$ 1--0 and 3--2 images the line
demarcating the red and blue emission is curved. Since it is seen in
both transitions, it is likely real and associated with
self-absorption in the presence of inward motions (see \S
\ref{s:l1489}). The position angle of $65^\circ$ agrees closely with
orientation of the equatorial plane inferred from near-infrared image
obtained with NICMOS on HST that reveals a monopolar reflection nebula
perpendicular to a dark lane \citep{padgett:nicmos}. The elongated
structure coincident with the dark lane is clearly seen in HCO$^+$
1--0 and 3--2, HCN 1--0, and $^{13}$CO 1--0, and at lower
signal-to-noise in C$^{18}$O 1--0. In addition, HCO$^+$ and HCN 1--0
trace extended emission near the source's systemic velocity to the
northeast of L1489~IRS. Figure \ref{f:l1489_mosaic} shows the image
obtained from a three-point mosaic with BIMA in C-array, covering
L1489~IRS and the starless core to its east, first identified by
\citet{mrh:scuba} as L1489~NE-SMM. This core does not show up
prominently in HCO$^+$ 1--0, but its southern half is traced by HCN,
indicating chemical differentiation along the starless core and
between this core and the material around L1489~IRS. In one possible
scenario, HCO$^+$ traces the densest material close to the star, while
HCN, which freezes out onto dust grains more easily, traces the warm
material close to the star and the southern, chemically `younger' part
of the starless core. Chemical model calculations
\citep{bergin:chem97} indicate that HCN is more abundant than HCO$^+$
at early times ($<10^6$~yr or, equivalently, lower densities), after
which it depletes rapidly while HCO$^+$ increases.

Chemical differences are also present in the emission toward TMC~1
(Fig. \ref{f:vcen}). In HCO$^+$ 1--0, $^{13}$CO 1--0, and
C$^{18}$O 1--0 an elongated core can again be seen around the YSO, at
a position angle of $90^\circ$. This is coincident with the equatorial
plane as inferred from single-dish CO 3--2 maps of its bipolar outflow
and scattered-light near-infrared imaging \citep{mrh:taurus2}; these
observations also suggest an inclination close to $90^\circ$
(=edge-on). HCN does not prominently trace this core, but, instead,
shows up strongly along the northern, blue-shifted outflow lobe. The
ratio of the three hyperfine components indicates an a moderate
opacity of 2--3 in the main component's line at the peak of the
emission. The images in all three hyperfine components are very
similar, however. A three-point mosaic obtained with in BIMA C-array
covering the outflow (Fig. \ref{f:tmc1_mosaic}) illustrates this
chemical differentiation more dramatically. HCN strongly peaks along
the northern flow, while HCO$^+$, although also revealing emission
along both lobes of the flow, has its maximum at the source
position. In fact, in the single pointing field of TMC~1 in  Fig.\
\ref{f:vcen}, HCO$^+$ emission associated with the outflow is
most prominent along the southern lobe, where no HCN emission is
detected. The appearance of TMC~1's continuum emission (Paper~I) rules
out a strong asymmetry in its envelope as an explanation for this
disparity. Similarly, in CO 3--2 the outflow appears symmetrical as
well \citep{mrh:taurus2}.  Differences in excitation conditions also
do not offer a likely explanation, since both HCN and HCO$^+$ 1--0
trace dense gas, with respective critical densities of $5\times 10^6$
and $2\times 10^5$~cm$^{-3}$. More likely, the HCN gas-phase abundance
is increased in the northern outflow lobe due to evaporation from icy
grain mantles. Subtle differences in shock conditions between the two
lobes appears to have lead to differences in temperature and
evaporation rate. Toward a sample of high-mass YSOs,
\citet{lahuis:iso_c2h2_hcn} find that the HCN gas-phase abundance is
strongly dependent on gas kinetic temperature.  The emission of
HCO$^+$ and HCN associated with the outflow is very narrow in velocity
($< 0.5$ km~s$^{-1}$), indicating that, since the system is seen close
to edge-on, any shocks driven into the material by the flow have a
very small transverse component.

A narrow velocity extent is a general characteristic of the emission
associated with TMC~1, as is immediately obvious from the comparison
of Fig.\ \ref{f:vcen} which are plotted
using the same range for the color scaling. Close inspection also
reveals that, although both sources appear similar in that they are
embedded in elongated cores, these cores might be rather different in
nature: The core around TMC~1 does not reveal the ordered red/blue
structure seen toward L1489~IRS, although a small gradient along the
equatorial plane is present. The core around TMC~1 extends further, by
a factor of 2--3 from the YSO than the one around L1489~IRS. Finally,
emission from the core around TMC~1 is predominantly located to one
side of the source (east), with only the C$^{18}$O emission symmetric
around the YSO. This lopsidedness is particularly clear in $^{13}$CO,
where the emission appears to peak $10''$ east of the source position
as inferred from continuum emission (\S \ref{s:cont} and Paper~I). A
second emission peak in $^{13}$CO is present $25''$ west of the
source, however, and \citet{mrh:taurus2} find a remarkably large
opacity ($\tau({\rm ^{13}CO}) \gg 3$) from single-dish $^{13}$CO and
C$^{18}$O 3--2 data. This leads us to suggest that significant
fractions of the emission in HCO$^+$, HCN, and $^{13}$CO may be
obscured by resolved-out foreground material.

The velocity structure of both YSOs is presented in Fig.\ \ref{f:pv}
through position-velocity diagrams taken along the equatorial planes
of the objects. L1489~IRS, in HCO$^+$, $^{13}$CO, and, at lower
signal-to-noise, in C$^{18}$O, shows the classical signature of
rotation with emission in the first and third quadrants and largest
velocities close to the source. The velocity structure of HCN includes
material near the systemic velocity associated with the starless core
east of the YSO. The nature of the velocity field of TMC~1 is far less
clear. \citet{mrh:taurus2} and \citet{brown:rotenv} interpreted the
position-velocity diagram of TMC~1, based on lower sensitivity HCO$^+$
data, as due to Keplerian rotation. In Fig. \ref{f:pv}, the lower-left
quadrant of TMC~1's HCO$^+$ diagram is indeed superficially similar to
L1489~IRS, and can be fit with rotation around a 0.8 $M_\odot$ object
(but see \S \ref{s:tmc1}). The corresponding emission in the first
quadrant is missing, and both authors note that this source is heavily
obscured, invoking absorption of the red-shifted HCO$^+$ emission as
the most likely cause. Unlike L1489~IRS, however, TMC~1's HCO$^+$
emission substantially `spills over' into the second quadrant, and its
C$^{18}$O emission is symmetric around the source center and the
systemic velocity. Together with the spatial discrepancies noted in
the previous paragraph, this leads us to suggest that infall combined
with heavy obscuration is a better description of TMC~1 than is
rotation alone. Section \ref{s:tmc1} explores this suggestion
quantitatively.

\section{Models for the emission\label{s:models}}

This section models the density and velocity distribution of the
material around L1489~IRS and TMC~1, trying to ascertain to what level
rotation and infall shape the emission. The best-fit models derived
from the SCUBA observations of Paper~I serve as starting points. In
particular, we adopt an outer radius of 2000~AU and a mass of
0.02~$M_\odot$ for the envelope around L1489~IRS, and the parameters
of the inside-out collapse model for TMC~1: a sound speed $a$ of
0.19~km~s$^{-1}$ and an age $t$ of $5.4\times 10^5$~yr, with an
envelope radius of 12,000 AU and containing 0.59~M$_\odot$. The model
calculations presented in this section use the two-dimensional
molecular excitation and radiative transfer Monte-Carlo code developed
by \citet{mrh:code}.

\subsection{A model for L1489~IRS\label{s:l1489}}

The velocity structure of L1489~IRS's envelope is clearly dominated by
rotation. Infall alone produces a position-velocity diagram where the
emission is distributed symmetrically around the source center and
systemic velocity; only self-absorption may introduce a measure of
asymmetry around the systemic velocity but not the source center (the
well known red/blue asymmetry of infall line profiles, for
example). This section investigates to what extent the emission around
L1489~IRS can be fit with rotation alone, or if infalling motions are
also required.

We start with a simple model for a circumstellar disk. The surface
density is given by
\begin{equation}
\Sigma = \Sigma_0 (R/1000\,{\rm AU})^p,\label{e:sigma}
\end{equation}
where $\Sigma_0$ is the density at a characteristic radius of 1000~AU
and $p$ is a power-law index. We consider a constant density model
with $p=0$, and a model with outward decreasing density, $p=-1.5$, as
invoked in many theoretical disk models and models of the primitive
solar nebula
\citep[e.g.]{chiang:disksed,lyndenbell:disks,weidenschilling:solarnebula}.
$\Sigma_0$ is chosen such that the observed dust mass of
0.02~$M_\odot$ (Paper~I) is contained in the model, for an outer
radius of 2000 AU and a gas-to-dust mass ratio of 100:1. For $p=0$,
$\Sigma_0=0.14$ kg~m$^{-2}$; for $p=-1.5$, $\Sigma_0 = 3.24$
kg~m$^{-2}$.

In addition to thin disk models, we also consider flared models, with
a scale height $h$ which increases linearly with radius, $h=R/2$. In
reality the value of the scale height will depend on the thermal
structure of the disk, an aspect that we neglect here. Instead we use
a maximum scale height of 1000 AU, which is the largest value
consistent with the observed aspect ratio of 2:1 and a maximum
inclination of $90^\circ$ (i.e., orienting the disk such that it
appears as slim as possible). A flat disk needs to be viewed under an
inclination of $60^\circ$ to produce the same aspect ratio. Scattered
light near-infrared imaging \citep{padgett:nicmos} and modeling of the
infrared spectral energy distribution \citep{kenyon:sed_2} suggest
source inclinations in this range of $60^\circ$--$90^\circ$ for
L1489~IRS, and we adopt $60^\circ$ and $90^\circ$ for the thin and
flared disk models, respectively.

The calculations use the H$_2$ volume density rather than a surface
density. For the thin disks, the material is distributed over a 10~AU
layer, much smaller than the disk's extent. For the flared models we
use
\begin{equation}
n(R,z) = \Sigma(R) / (2h(R) \sqrt{\pi}) \,
{\rm e}^{-z^2/h^2},\label{e:density}
\end{equation}
where $\Sigma(R)$ follows Eq. (\ref{e:sigma}) with $p=0$ or $p=-1.5$.

The velocity field in the disk follows Keplerian rotation around a
central mass, $V_\phi = \sqrt{G M_\star / R}$, with the central mass
$M_\star$ a free parameter, and no vertical motions, $V_z=0$. Inward
motions are parameterized by $V_R = -V_{\rm in} (R/100\,{\rm AU})^s$,
where $V_{\rm in}$ is a free parameter and $s=0$ (constant velocity)
or $s=-0.5$ (appropriate for free-fall). A turbulent line width of 0.2
km~s$^{-1}$ (FWHM) is also included, which is much smaller than the
resulting lines and does not affect the results significantly.

The final ingredient of the model is the kinetic temperature. We adopt
$T(R) = 34 (R/1000\,{\rm AU})^{-0.4}$~K which successfully fitted the
SCUBA observations (Paper~I). Since the excitation of HCO$^+$, the
molecule we will focus our attention on, is only modestly dependent on
kinetic temperature compared to density, the adopted kinetic
temperature is not very critical.

The model presented has two free parameters, $M_\star$ and $V_{\rm
in}$, and three additional parameters: thin or flared; $p=0$ or
$-1.5$; and $s=0$ or $-0.5$. Figure \ref{f:chi} shows the reduced
$\chi^2$ surfaces of a full parameter study in $M_\star$ and $V_{\rm
in}$ for the eight combinations of the three additional
parameters. The $\chi^2$ values are defined as
\begin{equation}
\chi^2 = {1\over N} \, \Sigma \Bigl({{T_{\rm model} - T_{\rm
obs}}\over{\sigma_{\rm obs}}}\Bigr)^2,\label{e:chi}
\end{equation}
where the summation is over all $N$ pixels along the equator and
spectral channels where HCO$^+$ 1--0 emission is detected in the
position-velocity diagram. Optically thin non-LTE excitation was
adopted, and the emission was scaled to the maximum intensity of the
observed image. This approximation decreased the required computing
time considerably, without significantly changing the results: the
$\chi^2$ value of the best-fit model changes by less than 40\% if
radiative trapping is included. This means that radiative coupling and
opacity are only important as a second-order effect for the overall
match between models and observations since the $\chi^2$ is dominated
by emission in the line wings which are optically thin.

The best fits, as judged from the minimum values in the $\chi^2$
surfaces, are obtained for models with a power-law distribution in
surface density ($p=-1.5$), without a clear preference for thin or
flared disks. The allowed range in $V_{\rm in}$ is significantly
larger for models with an free-fall-type velocity field ($s=-0.5$)
than for constant-velocity models. This is a direct result from the
power-law behavior of the former, which sets up a large range of
infall velocities over the extent of the envelope, decreasing the
sensitivity to the exact value of $V_{\rm in}$ at a particular
radius. Still, static models with $V_{\rm in}=0$ km~s$^{-1}$ can be
excluded regardless of the value of $s$ because the minimum $\chi^2$
values always lie above the $V_{\rm in}=0$ axis. Typical minimum
$\chi^2$ values are 3 for the models with $p=-1.5$ and 6--7 for
$p=0$. These low $\chi^2$ values indicate that the HCO$^+$
observations of L1489~IRS are well described by a 2000~AU radius disk,
in Keplerian rotation with infalling motions, and with a power-law
density distribution. The mass of the central object is $M_\star =
0.65$ $M_\odot$; and infall speeds are $V_{\rm in} = 1.3 \pm 0.2$
km~s$^{-1}$ for $s=-0.5$. Such models also reproduce the continuum
emission distribution as observed by SCUBA. The separation of the
velocity field in a Keplerian component and a radial component can
also be thought of as sub-Keplerian motions around an object exceeding
0.65~$M_\odot$ with material slowly spiraling inward. If the inward
motions correspond to free-fall, a central mass of 0.75~M$_\odot$ is
found.  The object's bolometric luminosity of 3.7~L$_\odot$ places an
upper limit of 1.4~M$_\odot$ on its mass, if the object is assumed to
be on the zero-age main-sequence and all luminosity is stellar
\citep{allens:p395}.

From observed H$^{13}$CO$^+$ single-dish lines \citep{mrh:taurus1}, an
average HCO$^+$ abundance is found of $(8 \pm 2)\times 10^{-9}$ with
respect to H$_2$.  Figure \ref{f:l1489_bestfit} shows the resulting
position-velocity diagrams for thin and flared disk models, now
including full non-LTE excitation and self-consistent two-dimensional
radiative transfer, and using $s=-0.5$. From the match of the emission
peaks, this figure suggests that a flared model is a slightly better
fit than the thin disk model. The $\chi^2$ method described above did
not distinguish between the two, since it represented a goodness of
fit over the entire emission region, yielding $\chi^2=2.8$ for
both. Including line trapping increases the $\chi^2$ values of the
flared model to 3.9 and of the flat model to 4.0. The flared model
also better fits the HCO$^+$ 3--2 position-velocity diagram; these data
were not included in the $\chi^2$ solution. The predicted
single-dish HCO$^+$ 1--0, 3--2, and 4--3 lines, as well as the HCO$^+$
3--2 spectrum in the interferometer beam lend further support to the
flared model (Fig.\ \ref{f:spectra}), especially the observed
intensities, widths, and profile shapes of the 3--2 and 4--3
lines. Only this model has lower excitation (lower density) material
in front of, and absorbing against, higher excitation gas, because of
both its exponential vertical distribution and its edge-on
orientation. The same effect is reflected in curvature of the line
separating red-shifted and blue-shifted HCO$^+$ 1--0 and 3--2 emission
around L1489~IRS in Fig. \ref{f:vcen}.  This line profile asymmetry also shows
that the material is moving inward and not flowing outward.

The profiles of the HCO$^+$ 1--0, and to a lesser extent, 3--2 lines
reveal discrepancies near the systemic velocity between the
observations and model, however. The single-dish data show additional
emission here, while the intensities are overestimated in the
interferometer beams. A likely explanation is offered by foreground
material associated with the overall cloud or the adjacent core
L1489~NE-SMM. Under fairly typical excitation conditions such
material can add significant flux in the lower HCO$^+$ lines, and
provide sufficient opacity to suppress the lines when it is
resolved-out by the interferometer beam. Adopting, somewhat
arbitrarily, a density of $n_{\rm H_2}=4\times 10^5$~cm$^{-3}$, and
kinetic temperature of $T_{\rm kin}=10$~K, a line width of $\Delta
V=1$~km~s$^{-1}$, and a HCO$^+$ column of $4\times
10^{12}$~cm$^{-2}$ for this putative foreground material, the observed
spectra are much better reproduced (see
Fig. \ref{f:spectra}). Additional emission is still required to match
the IRAM~30m data, but this may reflect material inside this $28''$
but not directly in front of L1489~IRS. \citet{mrh:thesis} and
\citet{girart:hco+ngc2264g} note that self-absorption in HCO$^+$ 1--0
can be severe toward the embedded YSOs L1527~IRS and NGC~2264G. We
suggest that such self-absorption in HCO$^+$ 1--0 is a general
characteristic of many embedded objects.

To summarize, L1489~IRS is a 0.65~$M_\star$ object embedded in a
rotating and infalling (or sub-Keplerian) disk, with a mass of
0.02~$M_\odot$, a radius of 2000~AU, at an inclination of
$60^\circ$--$90^\circ$, and likely flared. Absorption and emission in
HCO$^+$ 1--0 by unrelated foreground material is significant over a
limited velocity range. This is a common feature of HCO$^+$ 1--0
observations of embedded objects, and higher-excitation transitions of
HCO$^+$ are significantly less affected.

\subsection{A model for TMC~1\label{s:tmc1}}

The velocity structure in the envelope around TMC~1 has been described
previously \citep{mrh:taurus2,brown:rotenv} as Keplerian
rotation. This interpretation requires absorption by unrelated
foreground material to obscure the red-shifted emission. As the
previous section illustrated, such absorption is not
unlikely. However, Paper~I showed that the density distribution and
the velocity structure as probed through single-dish HCO$^+$ line
profiles is well described by the inside-out collapse model of
\citet{shu:selfsim}. Can these two description be reconciled?

Figure \ref{f:tmc1_mdl} compares the observed HCO$^+$
position-velocity diagram with two model predictions using the
collapse model of \citet{terebey:tsc}. This model describes the
self-similar, inside-out collapse of a cloud core with a small amount
of rotation, treating it as a perturbation to the solution of
\citet{shu:selfsim}. Since the sound-speed $a$ and age $t$ of the
model are already constrained by the SCUBA observations of Paper~I
($a=0.19$ km~s$^{-1}$; $t=2.5\times 10^5$ yr; $M_{\rm env}=0.59$
$M_\odot$), as is the HCO$^+$ abundance of $4\times 10^{-9}$ and the
temperature distribution of $T_{\rm kin} = 24\,(r/1000\,{\rm AU})^{-0.4}$~K,
the only free parameter is the rotation rate $\Omega$. The figure
shows results for $\Omega=0$ (no rotation) and $\Omega=3$
km~s$^{-1}$~pc$^{-1}$. The model does not include foreground
absorption, but the contours are dashed in the affected velocity range
by way of illustration.

The model results show that a collapse solution without rotation
already provides a reasonable fit to the data, and that $\Omega=3$
km~s$^{-1}$~pc$^{-1}$ is an especially close match. The resulting
azimuthal component is typical of those found toward embedded YSO,
e.g., L1527~IRS \citep{saito:l1551} and L1551~IRS~5
\citep{ohashi:l1527}. Rotating models also produce flattened cores
with the same elongation and offset from the source center in
integrated HCO$^+$ emission when taking foreground absorption into
account as observed (Fig. \ref{f:vcen}). A more formal $\chi^2$,
however, only weakly prefers $\Omega=3$ km~s$^{-1}$~pc$^{-1}$ over the
non-rotating model. A likely reason is that the model of
\citet{terebey:tsc} includes a perturbative treatment of
rotation. Material for which rotation dominates the velocity field is
not contained in the model; instead, all material within this
rotation-radius $R_c$ is assumed to collapse onto a circumstellar
disk. Emission from this disk component is not included in the model
calculations for lack of a continuous description throughout disk and
envelope. This means that the included rotational velocities do not
depend very sensitively on $\Omega$ once $R_c$ is of the order of a
few hundred AU, i.e., a significant fraction of the envelope.

Model calculations for $^{13}$CO and C$^{18}$O for $\Omega=3$
km~s$^{-1}$~pc$^{-1}$ also agree with the observed position-velocity
diagrams, using a CO depletion factor of 10--30 in material with
temperatures below 20~K, consistent with the findings of Paper~I. For
HCN, an upper limit to the abundance of $1\times 10^{-9}$ follows from
the lack of clear emission associated with the envelope. We conclude
that TMC~1 is embedded in an infalling envelope with some rotation:
$\Omega\approx 3$ km~s$^{-1}$~pc$^{-1}$, or equivalently, $R_c\approx
360$~AU. Within this radius $R_c$, material is expected to be in a
circumstellar disk.  The presence of rotation in the infalling
envelope is small enough that it does not significantly affect the
appearance of the line profiles in single-dish beams.

\subsection{Infall, rotation, and the evolution of young stellar
objects\label{s:evo}}

The results from the previous two sections indicate that infall and
rotation are present at different relative amounts in the envelopes
around L1489~IRS and TMC~1. Figure \ref{f:vfields} shows that infall
dominates the velocity field around TMC~1, while rotation dominates
around L1489~IRS. This, together with the smaller amount of
circumstellar material around L1489~IRS compared to TMC~1 (0.02
vs. 0.59 $M_\odot$, respectively), suggests that L1489~IRS is more
evolved than TMC~1. The small amount of rotation ($\Omega \approx
3$~km~s$^{-1}$~pc$^{-1}$) in TMC~1's infalling envelope has produced a
distinguishable flattening and rotational velocity component on scales
$\lesssim 3000$ AU, as probed by interferometric observations, but not
by single-dish spectra. At the source's age of $t=2.5\times 10^5$ yr
(for $a=0.19$ km~s$^{-1}$; Paper~I), this amount of initial rotation
has caused material inside $R_c=360$ AU to accrete onto a rotationally
supported disk. TMC~1's youth and its larger reservoir of
circumstellar material may be reflected in its more prominent outflow
compared to L1489~IRS. According to the adopted model
\citep{terebey:tsc}, the mass infall rate in the envelope is $\dot M
\approx 2\times 10^{-6}$ M$_\odot$~yr$^{-1}$. With a stellar mass of
$M_\star = \dot M \times t = 0.4$~M$_\odot$ and a stellar radius of
$R_\star \approx 3$~R$_\odot$, this gives an accretion luminosity of
$L_{\rm acc} = G M_\star \dot M / R_\star \approx 6$~L$_\odot$, much
larger than the measured bolometric luminosity of 0.7~L$_\odot$. Many
embedded YSOs suffer from this `luminosity problem'
\citep{kenyon:IRAS}, which is generally resolved by assuming that most
of the material accretes at larger radii onto a disk instead of
directly onto the star.

In this scenario, L1489~IRS is more evolved, and its disk has grown to
a radius of 2000~AU. In the model of \citet{terebey:tsc}, $R_c$
increases proportional to $t^3$, and it takes another 75\% of the
source's age (another $1$--$2\times 10^5$~yr; absolute time scales
depend on the sound speed) for $R_c$ to grow from TMC~1's value of
360~AU to 2000~AU. Over the same time span, the densities outside 2000
AU will have dropped below $\sim 10^4$~cm$^{-3}$ and the original
envelope is essentially depleted. These densities are now too low to
detect in, e.g., HCO$^+$ 1--0 emission, although they might contribute
to the absorption noted earlier. With the inferred inward motions, and
assuming that the disk will not acquire more material because of the
absence of extended emission in, e.g., the SCUBA observations of
Paper~I, in $2\times 10^4$~yr it will have contracted to a radius of
500~AU typical of disks around T~Tauri stars, much shorter than the
duration of the embedded phase of a few times $10^5$~yr. This time
scale is found by inverting the observed velocity field
$V(R)=dR/dt=V_{\rm in} (r/1000\,{\rm AU})^{-0.5}$, and integrating the
resulting expression for $dt/dR$ from 2000 to 500~AU.  Estimating the
mass accretion rate by dividing the disk's mass by its life time
yields $\dot M = 1\times 10^{-6}$~M$_\odot$~yr$^{-1}$ and $L_{\rm acc}
= 7$~L$_\odot$. This is comparable within a factor of two to the
measured bolometric luminosity of 3.7~L$_\odot$, suggesting that the
observed inward motions could correspond to accretion directly onto
the star itself. However, \citet{muzerolle:brgamma} infer a much lower
accretion luminosity of $\lesssim 0.3$ $L_\odot$ for L1489~IRS from
Br$\gamma$ observations. This indicates that most of the luminosity is
stellar, and that the inward motions in the 2000~AU disk give rise to
accretion on an object much larger than a star such as a circumstellar
disk.

Rotation and infall have been found in the envelopes around other
embedded YSOs, e.g., L1551~IRS~5 \citep{saito:l1551} and L1527~IRS
\citep{ohashi:l1527}. The infall motions toward these objects are
similar in magnitude as those for L1489~IRS and TMC~1, and rotational
velocities are smaller by factors of 3--10 compared to infall. This
suggests that they are very similar to TMC~1, that L1489~IRS's
velocity field stands apart, and that L1489~IRS represents a
short-lived, transitional stage between fully embedded objects and
T~Tauri stars with disks, equivalent to the last $\sim 5$\% of the
embedded phase. Given the number of embedded YSOs whose velocity
fields have been studied in detail \citep[so-far approximately
two dozen, e.g.]{mrh:scuba,chandler:scuba,shirley:scuba}, this is
consistent with L1489~IRS currently being the only representative of
its kind. Inspection of SCUBA data and HCO$^+$ 3--2 or 4--3 line
profiles may turn up more examples.
The velocity fields in the disks around
T~Tauri stars can be fit with Keplerian motion only
\citep{simon:ttsmass}, and the inferred accretion rates for T~Tauri
stars are much lower than that derived for L1489~IRS, of the order of
$10^{-8}$~M$_\odot$~yr$^{-1}$ \citep[e.g.]{calvet:ppiv}.

Understanding the formation of a disk from a collapsing envelope
requires a continuous theoretical description that encompasses the
transitional stage represented by L1489~IRS. Processes governing the
transport and dissipation of angular momentum will be important
ingredients of such theories. Theoretical descriptions are often
limited to times before or well after the formation of a rotating
disk. Some numerical simulations
\citep[e.g.]{yorke:disks3,nakamura:disklike} present results relevant
to objects like L1489~IRS. \citeauthor{yorke:disks3} find rotationally
supported structures of a few thousand AU radius as a natural outcome
of the collapse of a slowly rotating cloud core, but note that the
details of angular momentum transfer are critical.

There is no a priori reason to adopt an evolutionary interpretation
and to assume that TMC~1 will evolve to something resembling
L1489~IRS. However, non-evolutionary interpretations of the
differences between L1489~IRS's velocity structure and that of
embedded objects like TMC~1, L1527~IRS, and L1551~IRS, require some
explanation for its large rotational component. For example, L1489~IRS
may have originated from a core with an unusually large rotation
rate. Or mechanisms involved in carrying away excess angular momentum
like the stellar wind or the bipolar outflow may be frustrated.
However, the velocity gradient for the L1489 cloud on $1'$ scales was
measured at no more than 0.6~km~s$^{-1}$~pc$^{-1}$
\citep{goodman:cores8}, somewhat lower than the typical value for
dense cores of 1~km~s$^{-1}$~pc$^{-1}$. Also, near-infrared imaging
with NICMOS only revealed a single central source
\citep{padgett:nicmos}, ruling out that a binary companion with a
separation $>15$~AU has thwarted the growth of an accretion disk close
in around the star and thus shut down the likely driving mechanism of
the outflow and the star's mechanism for angular-momentum release
\citep[e.g.]{konigl:ppiv,shu:ppiv}. Since an evolutionary scenario
does not require that L1489~IRS is in some `special', we conclude that
it is a more appealing option. Identification of other objects
surrounded by thousand of AU scale disks is required to settle this
issue.

\section{Conclusion\label{s:conclusions}}

This paper presents interferometric molecular-line observations of two
embedded YSOs, L1489~IRS and TMC~1, and analyses the density and
velocity structure of their envelopes in terms of infall and
rotation. Our main conclusions are:

\begin{enumerate}
\item{L1489~IRS is surrounded by a 0.02~$M_\odot$, 2000~AU radius,
flared disk, which is in Keplerian rotation around a central object of
0.65~$M_\odot$ and which shows infalling motions. An equivalent
statement is that the disk shows sub-Keplerian motions around a
$\gtrsim 0.65$~$M_\odot$ but $\le 1.4$~M$_\odot$ star.  These will
reduce the size of the disk to the 500--800 AU typical of T~Tauri
stars in $2\times 10^4$~yr.}
  
\item{TMC~1 is embedded in a collapsing envelope of 0.58~$M_\odot$ and
12,000~AU in radius, which was originally rotating at a rate of
$\Omega =3$ km~s$^{-1}$~pc$^{-1}$, using the model of
\citet{terebey:tsc}. In another $2\times 10^5$~yr, the region
dominated by rotation, now limited to 360~AU, will have increased to
2000~AU.}
  
\item{We propose an evolutionary scenario, where TMC~1 represents the
earlier stage of the collapse of a cloud core with a small amount of
rotation. L1489~IRS is in a transitional stage, where the size of the
region dominated by rotation is at its maximum.  Subsequently, the
disk is expected to decrease in size, because the influx of fresh
material adding angular momentum to the system has ended, and the
stellar wind or other mechanisms continue to carry away angular
momentum. Equivalently, L1489~IRS's envelope can be described as a
sub-Keplerian, flattened structure slowly contracting to its
rotationally supported (Keplerian) size. A continuous description
which connects these stages continuously is clearly warranted.}
  
\item{Chemically, the observations indicate depletions by factors
10--30 for $^{13}$CO and C$^{18}$O by freezing out onto dust grains in
TMC~1's envelope, and abundances for HCO$^+$ of $8\times 10^{-9}$ in
the disk around L1489~IRS and $4\times 10^{-9}$ in the envelope around
TMC~1. In L1489~IRS's disk, HCN has an abundance of $(3\pm 1) \times
10^{-9}$, where temperatures may be high enough to evaporate part of
the ice mantles. An upper limit of $1\times 10^{-9}$ is found in
TMC~1's envelope, and the HCN abundance appears significantly enhanced
in the northern outflow lobe. Differences in HCO$^+$ and HCN emission
between the northern and southern outflow lobes indicate that subtle
changes in shock conditions bring about large differences in
evaporation rates.}

\end{enumerate}

\acknowledgments
The research of M.~R.~H. is supported by the Miller Institute for
Basic Research in Science. It is a pleasure to acknowledge the staffs
of the Owens Valley and Hat Creek observatories for excellent support
and hospitality during several observing runs. Ewine van Dishoeck and
Geoff Blake are acknowledged for careful reading of the
manuscript. The referee, Neal Evans, provided valuable comments that
improved the manuscript.




\newpage

\figcaption[fig1.eps]{Continuum emission at $\lambda=3.4$ and 2.7~mm
from L1489~IRS and TMC~1. {\it Left:\/} Cleaned images. The
synthesized beams are plotted in the lower left corners of each panel.
Contours are drawn at 3$\sigma$, 6$\sigma$, 9$\sigma$, $\ldots$; $\sigma=0.8$,
0.9, 1.3, and 0.8~mJy~beam$^{-1}$ from top to bottom.  {\it Right:\/}
Vector-averaged amplitudes as a function of $uv$-distance. The points
show the amplitudes in 10~k$\lambda$ wide bins, with 1$\sigma$ error
bars. The histograms indicate the expected value if no signal were
present in the data.\label{f:cont}}

\figcaption[fig2.eps]{Cleaned, integrated HCO$^+$ 1--0 emission of
L1489~IRS. {\it Left:\/} The image obtained using natural
weighting. {\it Right:\/} Same, for uniform weighting.  In both panels
contours are drawn using a square-root stretch to bring out the
extended emission with a factor of $\sqrt{2}$ increase per
level. Levels start at 0.048~Jy~beam$^{-1}$~km~s$^{-1}$ for the
naturally weighted image and at 0.070~Jy~beam$^{-1}$~km~s$^{-1}$ for
the uniformly weighted data.\label{f:l1489_robust}}

\figcaption[fig3.eps]{Integrated emission (contours) superposed on the
velocity centroid with respect to the sources' systemic velocity
(color) of HCO$^+$ 1--0 and 3--2 , HCN 1--0, $^{13}$CO 1--0, and
C$^{18}$O 1--0 toward L1489~IRS (top) and TMC~1 (bottom). The HCN
image includes the main hyperfine component only. Contours are drawn
using a square-root stretch to bring out the extended emission with a
factor of $\sqrt{2}$ increase per level, and start at 0.07, 3.1, 0.10,
0.08, and 0.06~Jy~beam$^{-1}$~km~s$^{-1}$ for HCO$^+$ 1--0, HCO$^+$
3--2, HCN 1--0, $^{13}$CO 1--0, and C$^{18}$O 1--0, for L1489~IRS
respectively. For TMC~1, they start at 0.04, 0.07, 0.05, and
0.05~Jy~beam$^{-1}$~km~s$^{-1}$ for HCO$^+$, HCN, $^{13}$CO, and
C$^{18}$O, respectively. The text describes the adopted weighting
schemes. The synthesized beams are plotted in the lower left corner of
each panel.\label{f:vcen}}


\figcaption[fig4.eps]{Image obtained from a three-point mosaic of
L1489~IRS in HCO$^+$ and HCN 1--0, covering the adjacent core
L1489~NE-SMM (Paper~I). The location and extent of this core are
indicated with the cross and the heavy dashed line. The HCN image
includes the main hyperfine component only. Contours and grey-scale
show integrated emission, with the contours drawn at 0.5, 1.0, 2.0,
4.0, $\ldots$ Jy~beam$^{-1}$~km~s$^{-1}$ for HCO$^+$ and 0.3, 0.6,
1.2, 2.4, $\ldots$ Jy~beam$^{-1}$~km~s$^{-1}$ for HCN. The thin dashed
line outlines the mosaicked region.\label{f:l1489_mosaic}}

\figcaption[fig5.eps]{Image obtained from a three-point mosaic of
TMC~1 in HCO$^+$ and HCN 1--0, covering the outflow. The HCN image
includes the main hyperfine component only. Contours and grey-scale
show integrated emission, with the contours drawn at 0.13, 0.26, 0.52,
1.0, $\ldots$ Jy~beam$^{-1}$~km~s$^{-1}$ for HCO$^+$ and HCN. The thin
dashed line outlines the mosaicked region.\label{f:tmc1_mosaic}}

\figcaption[fig6.eps]{Position-velocity diagrams obtained along the
equatorial plane of L1489~IRS ({\it left\/}) and TMC~1 ({\it right\/})
in the observed lines. Contours are drawn at linear intervals,
starting at 0.28 (HCO$^+$), 0.35 (HCN), 0.20 ($^{13}$CO), and
0.20~Jy~beam$^{-1}$ (C$^{18}$O) for L1489~IRS, and 0.15 (HCO$^+$),
0.25 (HCN), 0.20 ($^{13}$CO), and 0.20~Jy~beam$^{-1}$ (C$^{18}$O) for
TMC~1. The angular and velocity resolution is depicted by the filled
symbol in the lower left corner of each panel. The vertical and
horizontal lines indicate the sources' positions and systemic
velocities.\label{f:pv}}

\figcaption[fig7.eps]{Surfaces of reduced $\chi^2$ between the
observed HCO$^+$ 1--0 position-velocity diagram of L1489~IRS and the
model results. The free parameters of the model are the central mass
$M_\star$ and the infall speed $V_{\rm in}$; the additional parameters
of the model (thin or flared geometry; radial power-law index $s$ of
the velocity field; and radial power-law index $p$ of the surface
density) are indicated in each panel. Contours are drawn at $\chi^2$
intervals of 1. The minimum $\chi^2$ values are 6.1, 3.0, 5.9, and 2.8
(top panels, from left to right); 6.6, 2.8, 7.0, and 2.8 (bottom
panels, from left to right). The greyscale is the same for each panel
and is indicated at the lower right.\label{f:chi}}

\figcaption[fig8.eps]{HCO$^+$ 1--0 and 3--2 position-velocity diagrams
along the equator of the two best-fit models ({\it contours\/}; thin
and flared geometries) superposed on the observed position-velocity
diagram of L1489~IRS ({\it grey-scale\/}). Contour levels are the same
as for Fig.\ \ref{f:pv}. The first contour levels and the increase
between the levels are 0.28~Jy~beam$^{-1}$ and 3.1~Jy~beam$^{-1}$ for
1--0 and 3--2, respectively.\label{f:l1489_bestfit}}

\figcaption[fig9.eps]{Observed HCO$^+$ spectra ({\it histograms\/}) of
L1489~IRS in the interferometer beams (1--0 and 3--2), the $28''$
IRAM~30m beam (1--0), and $19''$ and the $14''$ JCMT beams (3--2 and
4--3). The line profiles produced by the thin ({\it dashed curve\/})
and flared ({\it solid curve\/}) disk models are superposed. The
excess HCO$^+$ 1--0 emission around a $V_{\rm LSR}$ of 7 kms~s$^{-1}$
likely originates in unassociated material along the line of sight. A
possible fit to this foreground material obscuring the flared-disk
model is shown in the heavy solid curves, adopting $n_{\rm
H_2}=4\times 10^5$~cm$^{-3}$, $T_{\rm kin}=10$~K, $\Delta
V=1$~km~s$^{-1}$, and a HCO$^+$ column of $4\times
10^{12}$~cm$^{-2}$.\label{f:spectra}}

\figcaption[fig10.eps]{Position-velocity diagram of HCO$^+$ 1--0
observed toward TMC~1 ({\it grey-scale\/}) over-plotted with model
predictions based on the inside-out collapse model of Paper~I. {\it
Left\/}: no rotation; {\it right\/}: slow rotation,
$\Omega=3$~km~s$^{-1}$~pc$^{-1}$. The model results are sampled at the
same $(u,v)$ positions as the data are deconvolved in a similar way,
and hence accurately account for the resolving out of larger scale
emission. Model emission in the $V_{\rm LSR}$ range of
5.4--7.0~km~s$^{-1}$ is plotted with dashed lines to reflect the
obscuration by unrelated foreground material.\label{f:tmc1_mdl}}

\figcaption[fig11.eps]{Comparison of the radial (`infall') and
azimuthal (`rotation') components of the velocity fields around
L1489~IRS and TMC~1. \label{f:vfields}}


\newpage


\begin{deluxetable}{lll}
\tablecaption{Observations\label{t:obs}}
\tablecolumns{3}
\tablehead{\colhead{Date} & \colhead{Instrument}}
\startdata
1993/10; 1994/2--4 & OVRO, L+E & HCO$^+$ 1--0, 
  3.4~mm continuum\tablenotemark{a}\\
1998/10,11; 1999/2--4 & BIMA, B+C & HCO$^+$ 1--0, HCN 1--0, 3.4~mm
  continuum\\
1995/2--5; 1996/10,11; 1997/2 & OVRO, L+E & $^{13}$CO, C$^{18}$O 1--0,
   2.7~mm continuum\tablenotemark{a}\\
1999/11,12; 2000/1--2 & OVRO, L+H & $^{13}$CO, C$^{18}$O 1--0,
   2.7~mm continuum\\
2000/12 & BIMA, C & HCO$^+$ 3--2, 1.1~mm continuum
\enddata
\tablenotetext{a}{Previously presented in \citet{mrh:taurus1,mrh:taurus2}.}
\end{deluxetable}

\begin{deluxetable}{lrrrrr}
\tablecaption{Continuum emission\label{t:cont}}
\tablecolumns{6}
\tablehead{ & \colhead{$\alpha$(2000.0)\tablenotemark{a}} & 
  \colhead{$\delta$(2000.0)\tablenotemark{a}} & 
  \colhead{$\lambda$} & \colhead{Flux} & \colhead{Beam Size}\\
\colhead{Source} & \colhead{(h m s)} & \colhead{($^\circ$ $'$ $''$)} &
  \colhead{(mm)} & \colhead{(mJy)} & \colhead{(arcsec)}}
\startdata
L1489~IRS & 04:04:42.96 & +26:18:57.1 & 3.4 & $6.4\pm 1.8$ & $5.7\times 4.8$\\
         &            &             & 2.7 &  $6.8\pm 1.3$ & $4.0\times 3.3$ \\
 & & & 1.1 & $<70$ & $4.6\times 2.5$ \\
TMC~1     & 04:41:12.69 & +25:46:35.2 & 3.4 & $9.5\pm 1.5$ & $4.4\times 4.1$\\
         &            &             & 2.7 & $10.3\pm 1.4$ & $3.0\times 2.6$ \\
\enddata
\tablenotetext{a}{Best-fit continuum position, adopted as source position.}
\end{deluxetable}

\begin{deluxetable}{llrrrr}
\tablecaption{Line emission\label{t:line}}
\tablehead{& & \colhead{Beam Size} & 
  \colhead{$\int T_b dV$\tablenotemark{a}} & 
  \colhead{$\langle \int T_b dV \rangle$\tablenotemark{b}} &
  \colhead{Region, Size}\\
\colhead{Source} & \colhead{Line} & \colhead{(arcsec)} & 
  \colhead{(K~km~s$^{-1}$)} & 
  \colhead{(K~km~s$^{-1}$)} & \colhead{(arcsec)}}
\startdata
L1489~IRS & HCO$^+$ 1--0\tablenotemark{c} & 
  $5.5\times 5.0$ & $ 32.9\pm 0.5$ & $ 6.1\pm 0.1$ & 
  Central core, $30\times 25$\\
 & HCO$^+$ 1--0\tablenotemark{d} & 
  $3.7\times 3.3$ & $ 38.6\pm 1.7$ & $ 7.8\pm 0.2$ & 
  Central core, $30\times 25$\\
 & HCO$^+$ 1--0\tablenotemark{e} & 
  $3.2\times 2.8$ & $ 33.0\pm 4.2$ & $ 9.6\pm 0.7$ & 
  Central core, $30\times 25$\\
 & HCN 1--0 & $6.1\times 4.9$ & $ 7.4\pm 0.7$ & $ 2.0\pm 0.2$ &
  Central core, $20\times 20$\\
 & $^{13}$CO 1--0 & $3.6\times 3.0$ & $ 23.2\pm 1.2$ & $ 5.5\pm 0.2$ &
  Central core, $20\times 20$\\
 & C$^{18}$O 1--0 & $5.8\times 4.7$ & $ 1.5\pm 0.3$ & $ 0.7\pm 0.1$ &
  Central core, $20\times 20$\\
 & HCO$^+$ 3--2 & $4.6\times 2.5$ & $77.9\pm 1.7$ & $1.9\pm 0.2$ & 
  Central core, $30\times 25$ \\
TMC~1 & HCO$^+$ 1--0 & $5.0\times 4.4$ & $ 10.2\pm 0.4$ & $ 2.4\pm 0.07$ &
  Central core, $25\times 20$\\
 & & & $ 10.2\pm 0.4$ & $ 1.7\pm 0.04$ &
  Entire field, $30\times 40$\\
 & & & $ 2.6\pm 0.4$ & $ 1.0\pm 0.08$ &
  Southern ridge, $20\times 20$\\
 & HCN 1--0 & $5.0\times 4.2$ & $ 2.8\pm 0.3$ & $ 1.2\pm 0.02$ &
  Outflow lobe, $50\times 65$\\
 & $^{13}$CO 1--0 & $4.9\times 4.5$ & $ 4.3\pm 0.2$ & $ 1.2\pm 0.1$ &
  Central core, $20\times 20$\\
 & & & $ 0.8\pm 0.2$ & $ 0.4\pm 0.1$ &
  Southern ridge, $10\times 20$\\
 & C$^{18}$O 1--0 & $4.9\times 4.5$ & $ 2.0\pm 0.4$ & $ 1.0\pm 0.2$ &
  Central core, $10\times 10$\\
\enddata
\tablenotetext{a}{Maximum integrated intensity over indicated region.}
\tablenotetext{b}{Average integrated intensity over indicated region.}
\tablenotetext{c}{Image reconstructed using natural weighting.}
\tablenotetext{d}{Image reconstructed using weighting with a
`robustness' parameter of 0.}
\tablenotetext{e}{Image reconstructed using uniform weighting.}
\tablenotetext{f}{Flux scaling uncertain because of atmospheric decorrelation.}
\end{deluxetable}



\newpage

\begin{figure}
\figurenum{\ref{f:cont}}
\epsscale{0.7}
\plotone{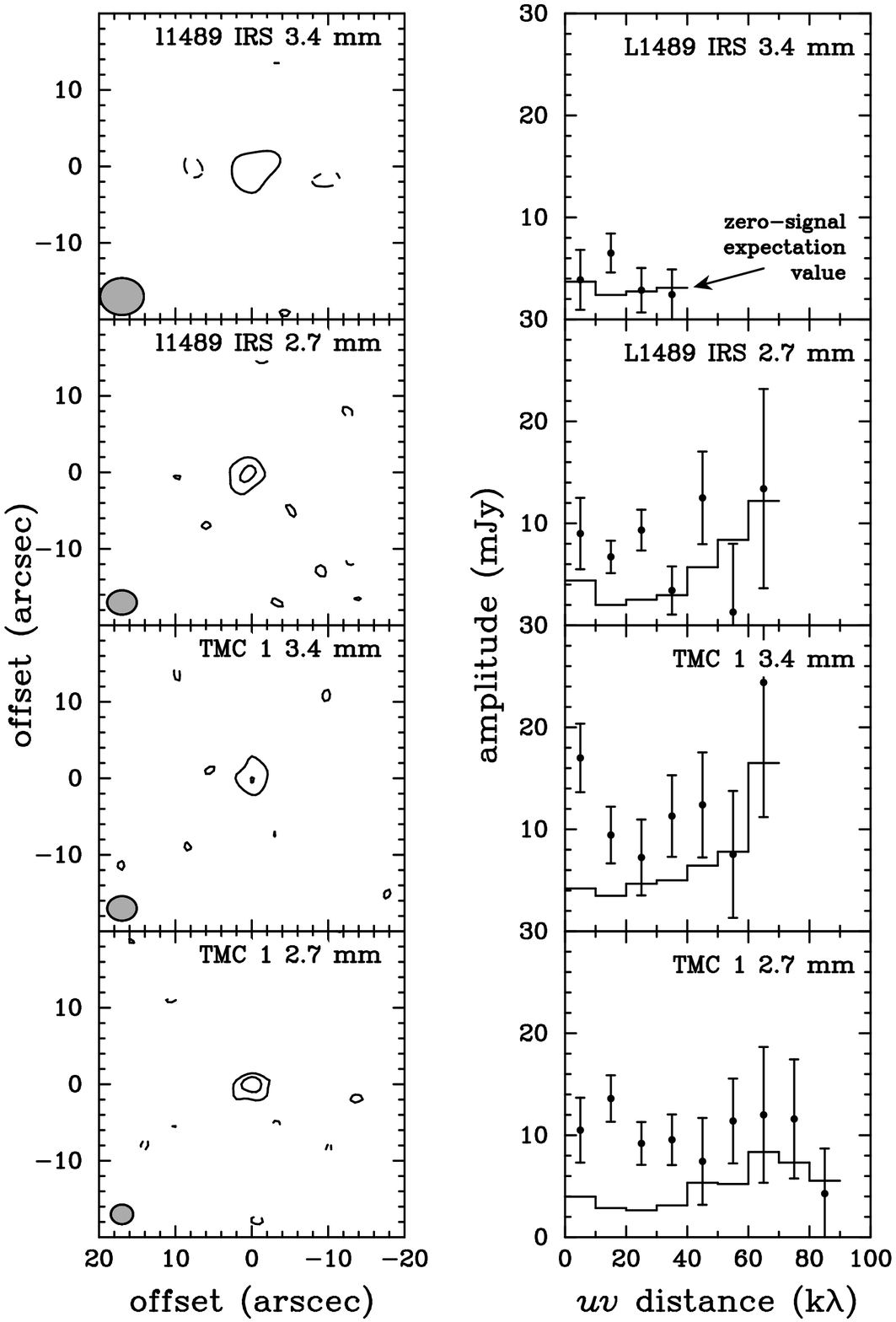}
\caption{}
\end{figure}

\begin{figure}
\figurenum{\ref{f:l1489_robust}}
\epsscale{1}
\plotone{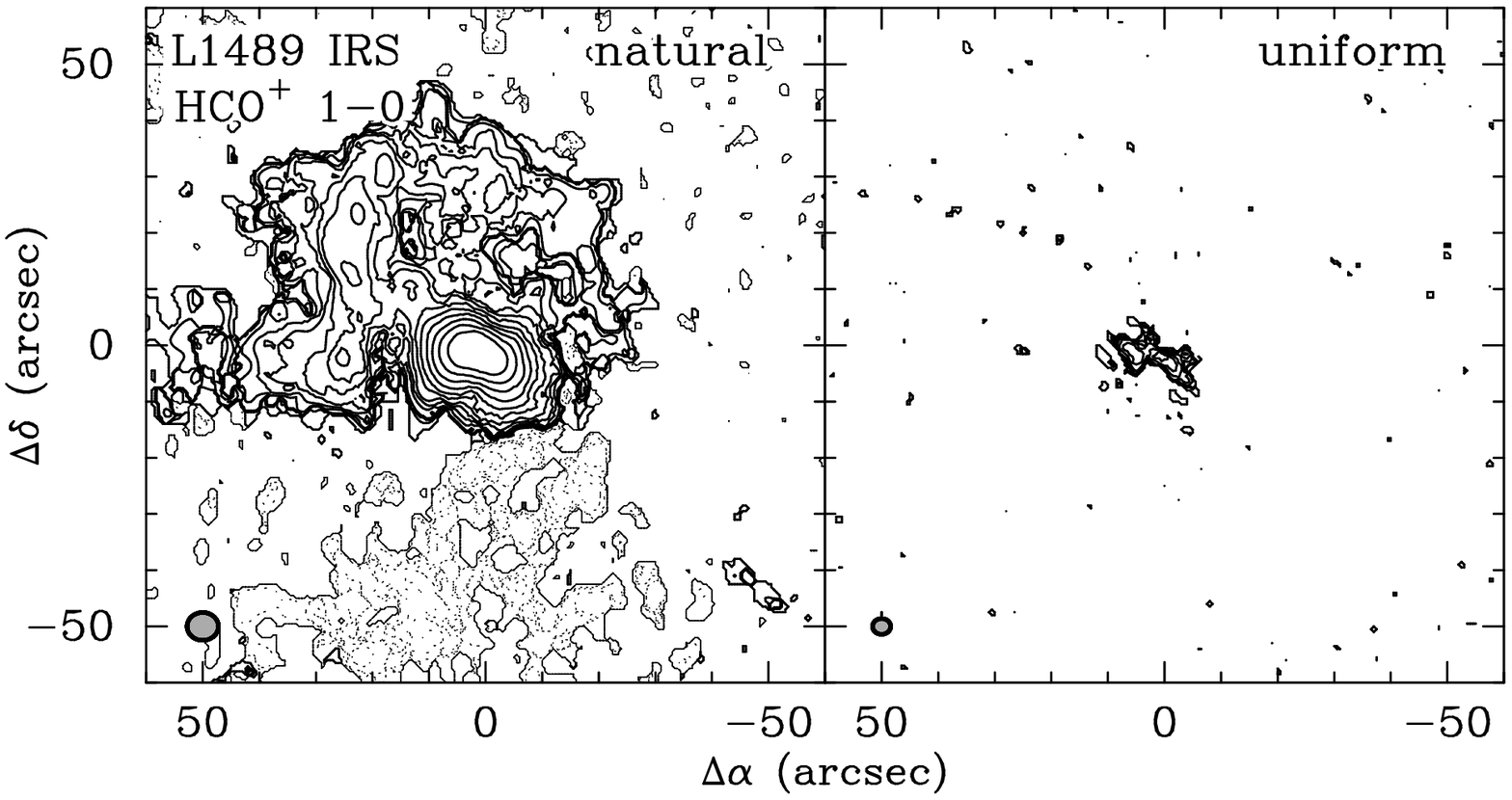}
\caption{}
\end{figure}

\begin{figure}
\figurenum{\ref{f:vcen}}
\epsscale{1}
\plotone{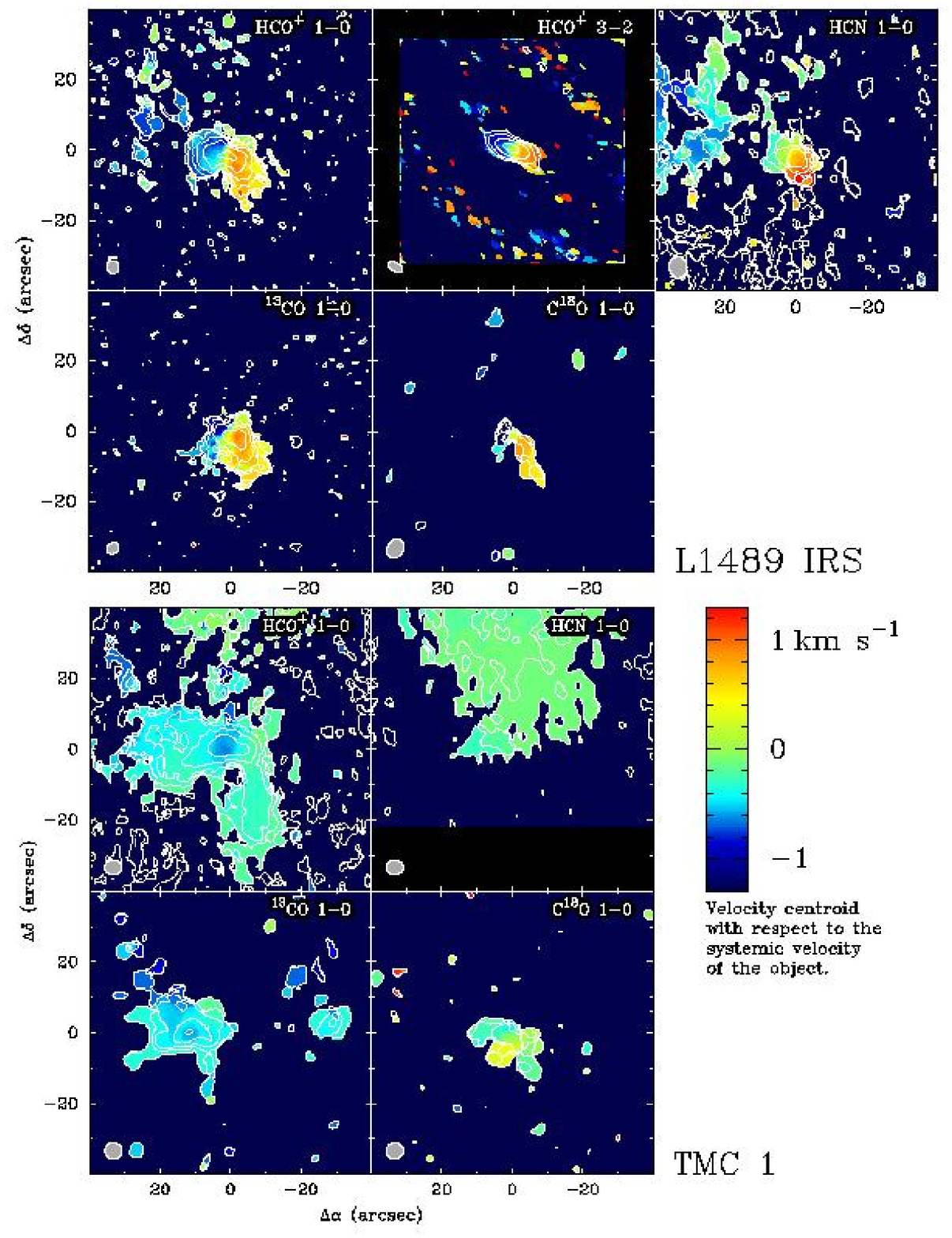}
\caption{}
\end{figure}


\begin{figure}
\figurenum{\ref{f:l1489_mosaic}}
\epsscale{1}
\plotone{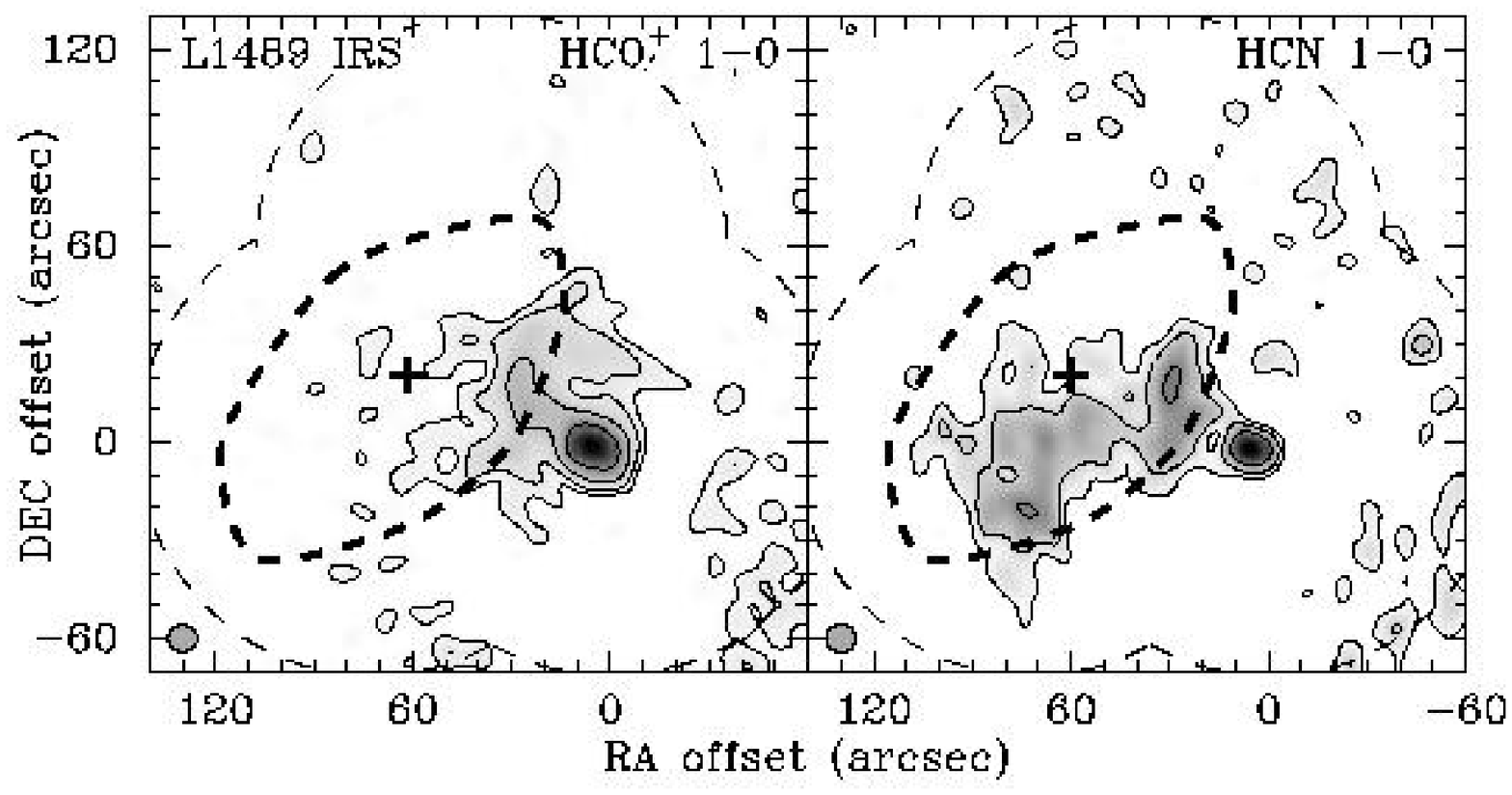}
\caption{}
\end{figure}

\begin{figure}
\figurenum{\ref{f:tmc1_mosaic}}
\epsscale{1}
\plotone{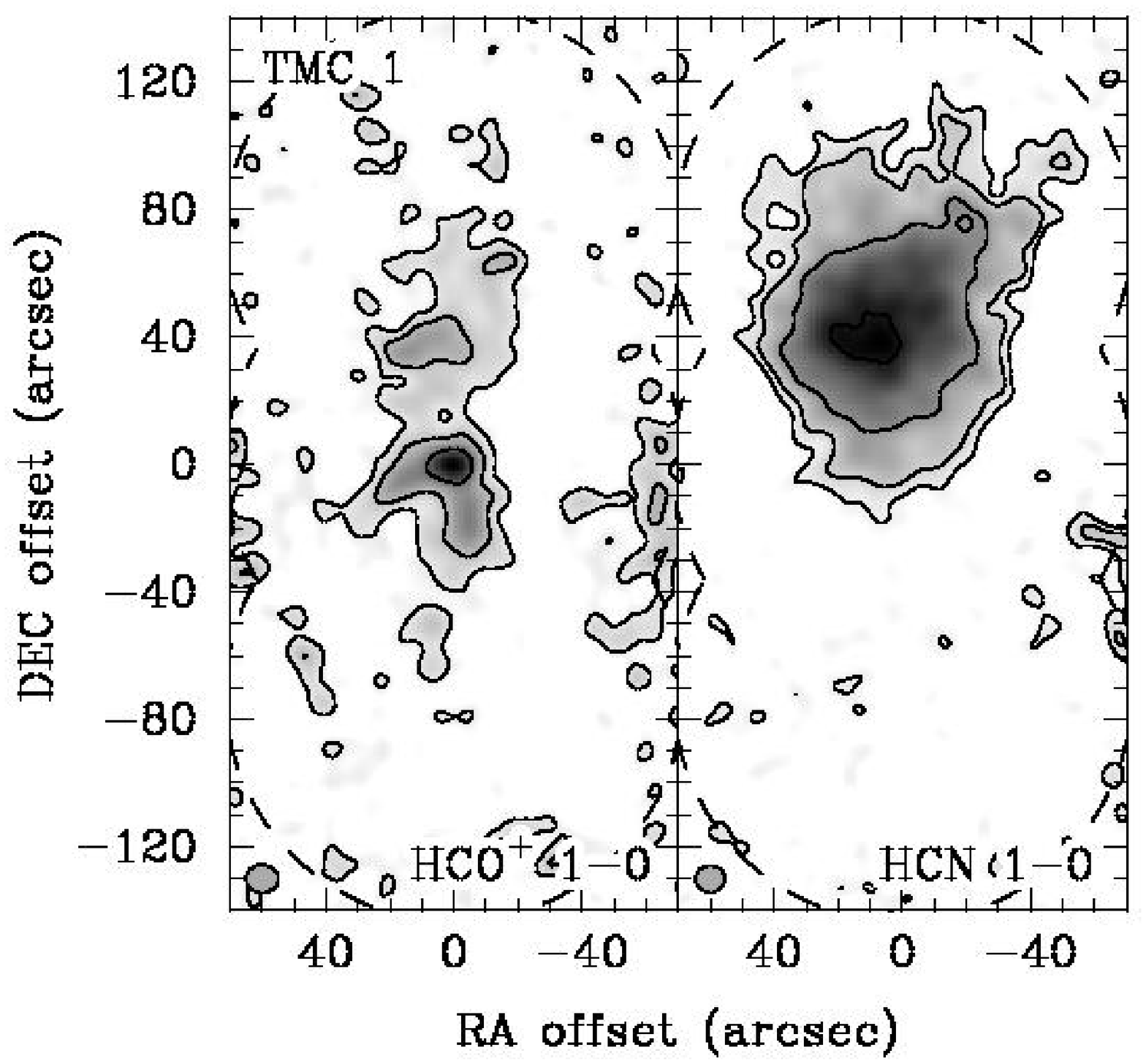}
\caption{}
\end{figure}

\begin{figure}
\figurenum{\ref{f:pv}}
\epsscale{0.5}
\plotone{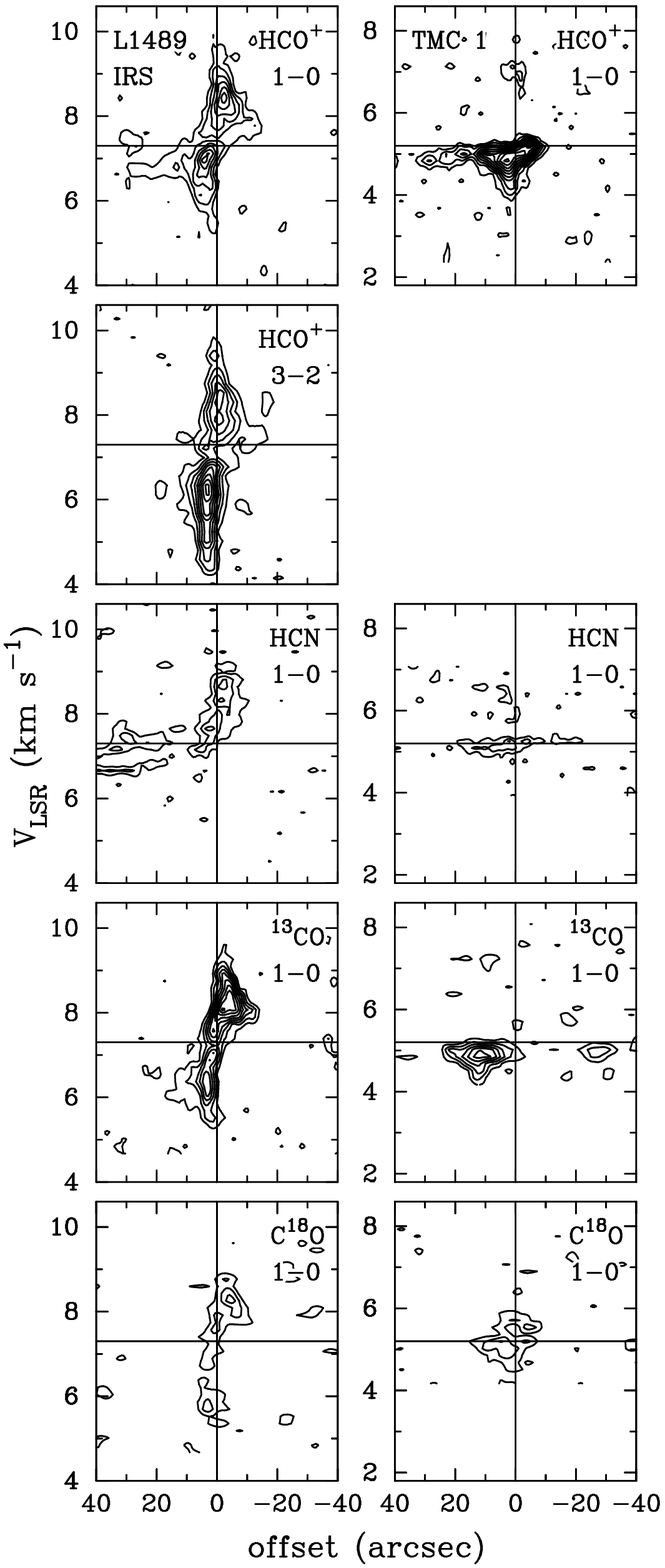}
\caption{}
\end{figure}

\begin{figure}
\figurenum{\ref{f:chi}}
\epsscale{1}
\plotone{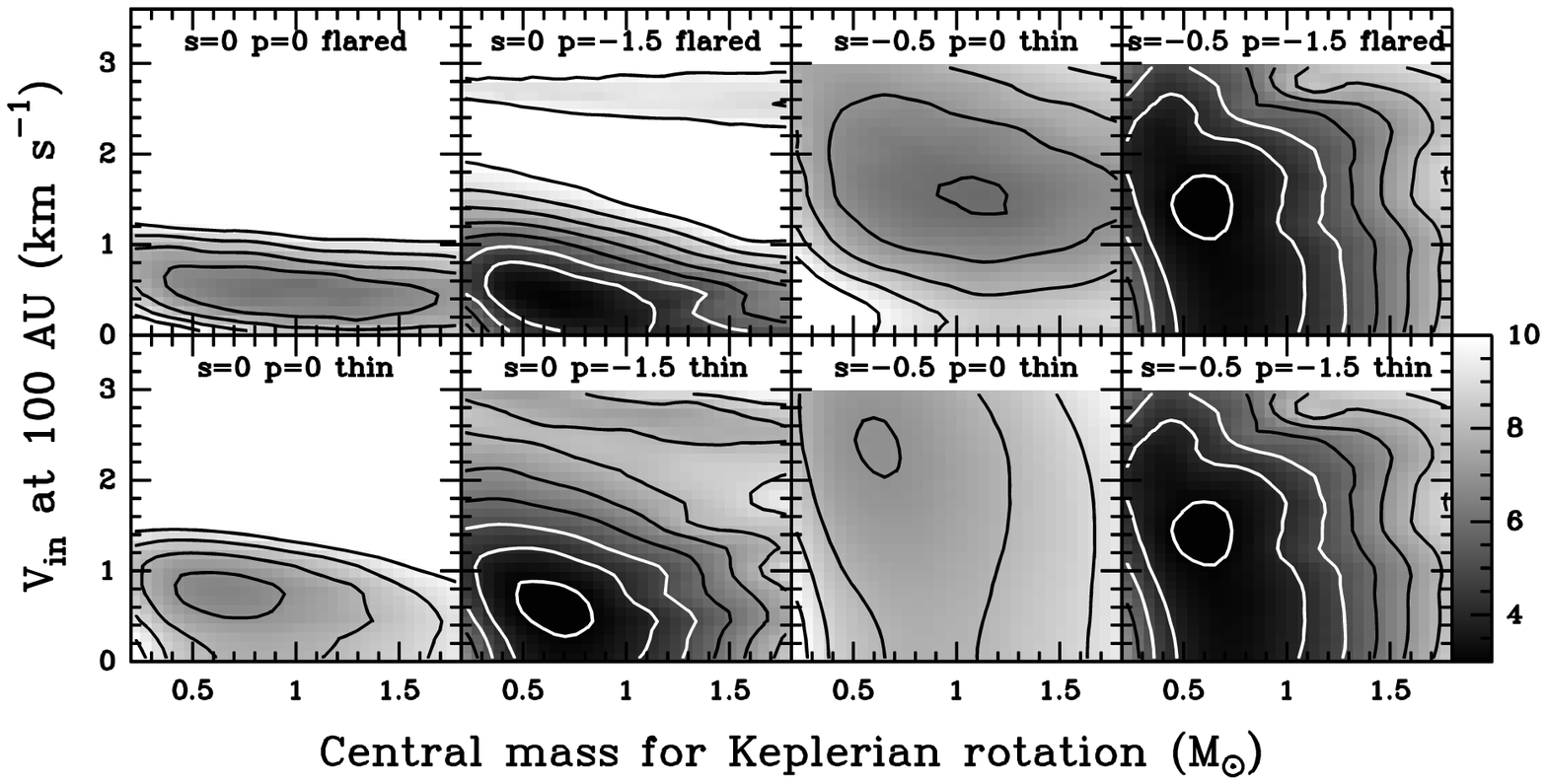}
\caption{}
\end{figure}

\begin{figure}
\figurenum{\ref{f:l1489_bestfit}}
\epsscale{1}
\plotone{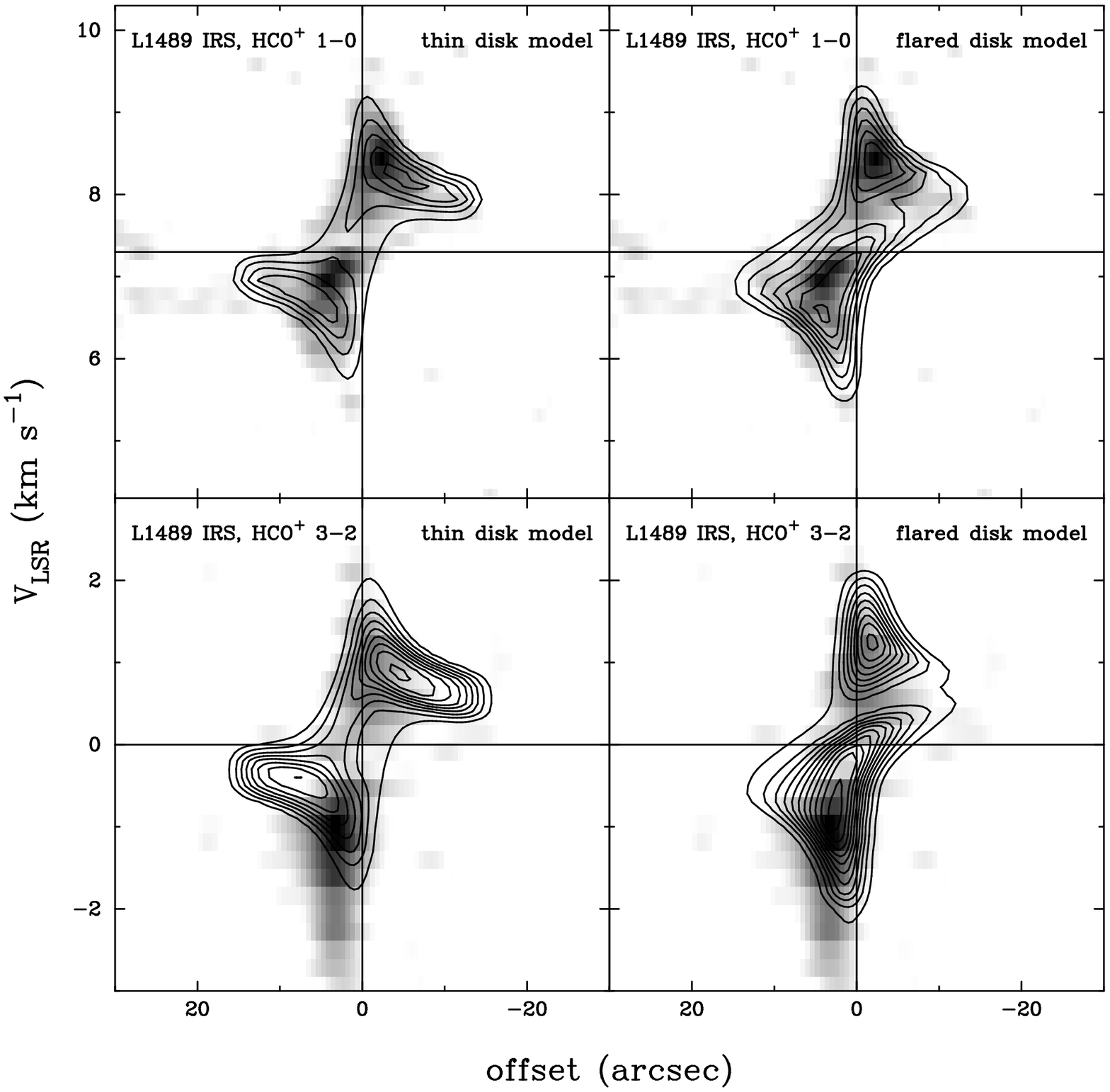}
\caption{}
\end{figure}

\begin{figure}
\figurenum{\ref{f:spectra}}
\epsscale{1}
\plotone{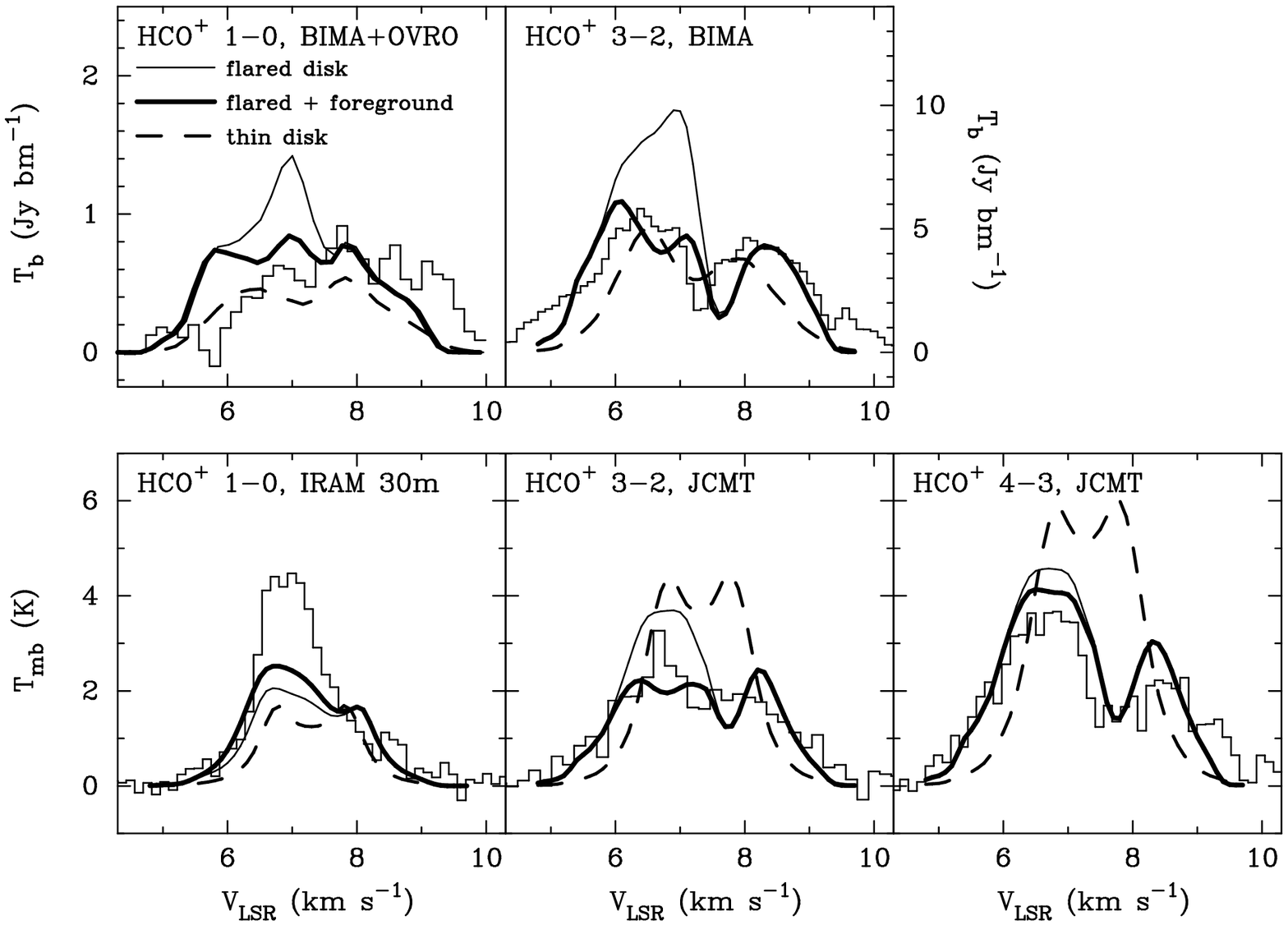}
\caption{}
\end{figure}

\begin{figure}
\figurenum{\ref{f:tmc1_mdl}}
\epsscale{1}
\plotone{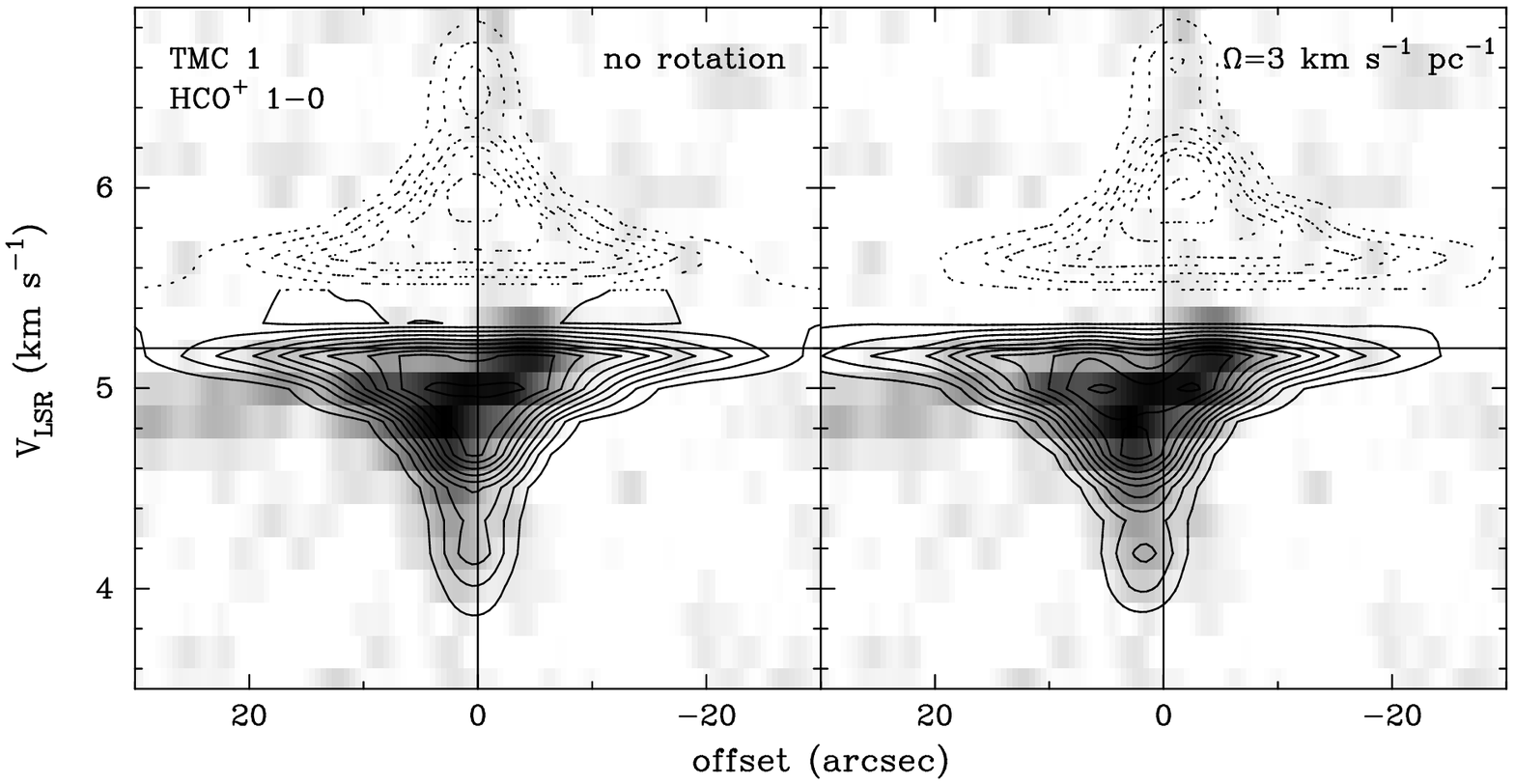}
\caption{}
\end{figure}

\begin{figure}
\figurenum{\ref{f:vfields}}
\epsscale{1}
\plotone{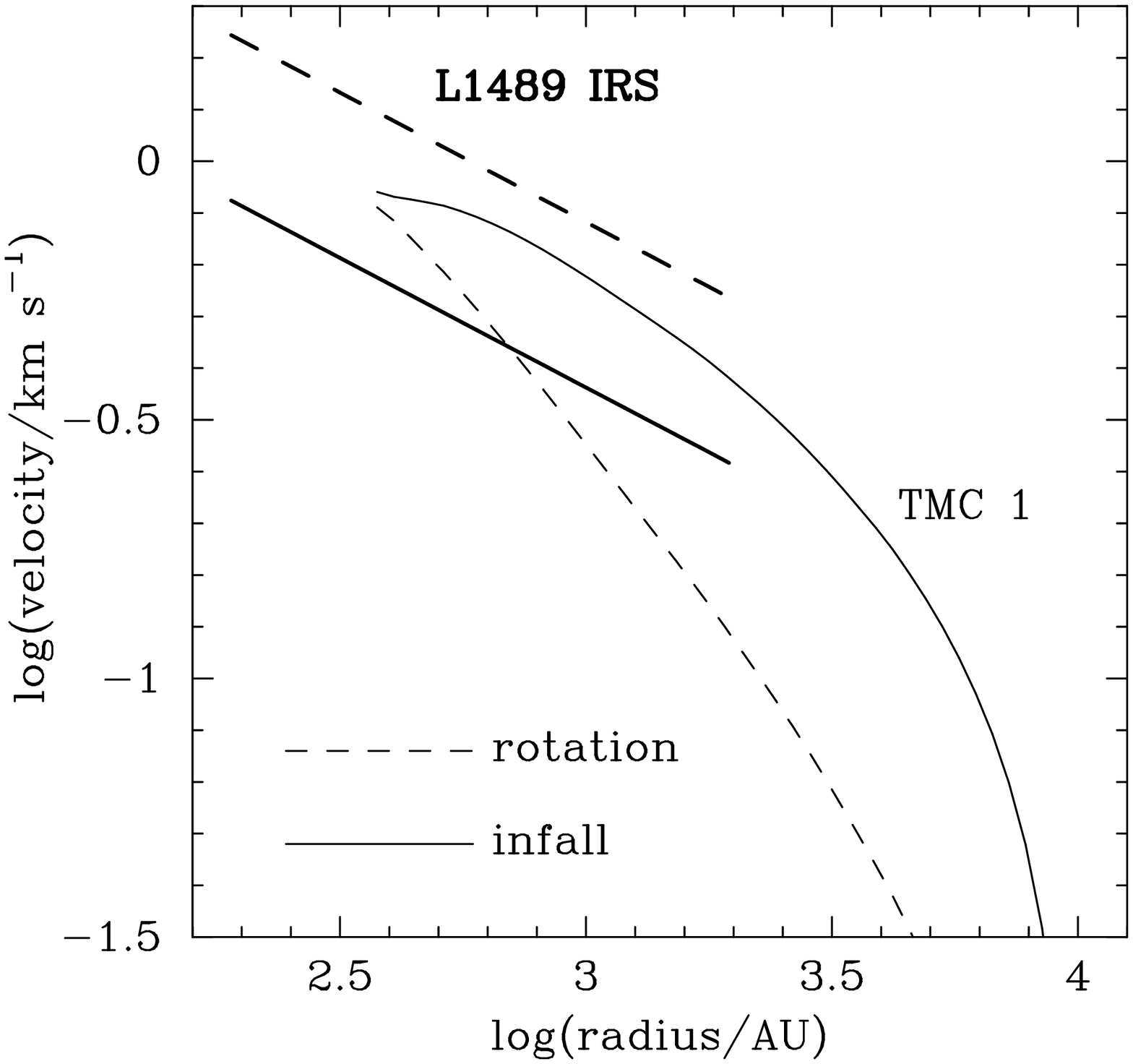}
\caption{}
\end{figure}


\end{document}